\documentclass[a4paper,11pt]{article}
\pdfoutput=1

\usepackage{jheppub}
\usepackage{pdflscape}
\usepackage{color}
\usepackage{tabularx}
\usepackage{hyperref}
\usepackage{amsmath}
\usepackage{amssymb}
\usepackage{graphicx,multirow,subfigure}
\usepackage[T1]{fontenc}
\usepackage{comment}
\usepackage[dvipsnames]{xcolor}
\usepackage{physics}
\usepackage{bbold}
\usepackage{mathtools}
\usepackage{cancel}

\renewcommand{\bar}{\overline}

\renewcommand \ket[1]{
        \left| #1 \right>
}

\renewcommand \bra[1]{
        \left< #1 \right|
}

\newcommand{\bi}{\begin{itemize}}
\newcommand{\ei}{\end{itemize}}
\newcommand{\ben}{\begin{enumerate}}
\newcommand{\een}{\end{enumerate}}
\newcounter{mycount}
\newcommand{\pauseen}{\setcounter{mycount}{\value{enumi}}\end{enumerate}}
\newcommand{\resumeen}{\begin{enumerate}\setcounter{enumi}{\value{mycount}}}

\newcommand{\SUt}{SU(3)$_{F}$}

\newcommand{\be}{\begin{equation}}
\newcommand{\ee}{\end{equation}}
\newcommand{\bea}{\begin{eqnarray}}
\newcommand{\eea}{\end{eqnarray}}
\newcommand{\beq}{\begin{equation}}
\newcommand{\eeq}{\end{equation}}
\def\beqa{\begin{eqnarray}}
  \def\eeqa{\end{eqnarray}}
\def\lsim{\mathrel{\rlap{\lower4pt\hbox{\hskip1pt$\sim$}}
    \raise1pt\hbox{$<$}}}         
\def\gsim{\mathrel{\rlap{\lower4pt\hbox{\hskip1pt$\sim$}}
    \raise1pt\hbox{$>$}}}         

\newcommand{\eq}[1]{Eq.~(\ref{#1})}
\newcommand{\eqsand}[2]{Eqs.~(\ref{#1}) and (\ref{#2})}

\begin{document}

\title{Anatomy of Non-Leptonic Two-Body Decays of Charmed Mesons into Final States with $\eta'$}

\author[a,b]{Carolina Bolognani}
\author[c]{Ulrich Nierste}
\author[d]{Stefan Schacht}
\author[a,b]{K. Keri Vos}

\affiliation[a]{Gravitational Waves and Fundamental Physics (GWFP), Maastricht University, Duboisdomein 30, NL-6229 GT Maastricht, the Netherlands}
\affiliation[b]{Nikhef, Science Park 105, NL-1098 XG Amsterdam, the Netherlands}
\affiliation[c]{Institute for Theoretical Particle Physics, Karlsruhe Institute of Technology (KIT), Wolfgang-Gaede-Str. 1, 76131 Karlsruhe, Germany}
\affiliation[d]{Department of Physics and Astronomy, University of Manchester, Manchester M13 9PL, United Kingdom}

\emailAdd{carolina.bolognani@cern.ch}
\emailAdd{ulrich.nierste@kit.edu}
\emailAdd{stefan.schacht@manchester.ac.uk}
\emailAdd{k.vos@maastrichtuniversity.nl}

\subheader{\hfill \textnormal{Nikhef 2024-017, TTP24-040}}

\abstract{
We show that $D\rightarrow P\eta^\prime$ decay amplitudes, where $P=K,\pi,\eta$, cannot be simply related to their $D\rightarrow P\eta$ counterparts with a single $\eta_0$--$\eta_8$ mixing angle. We propose a novel, consistent treatment of $\eta_0$--$\eta_8$ mixing for application in $D\to P\eta$ and $D\to P\eta^\prime$ decays. Using this framework, we perform a global analysis of $D\rightarrow P\eta^\prime$ decays employing SU(3)$_F$ symmetry including linear SU(3)$_F$ breaking. We find that the assumption of 30\% SU(3)$_F$ breaking is in slight tension ($2.5\sigma$) with the data when compared to a fit that allows for 50\% SU(3)$_F$ breaking, the latter giving a perfect description of the data. In order to allow for further scrutinization of SU(3)$_F$-breaking effects in the future, we give branching ratio predictions for all $D\rightarrow P\eta'$ modes. Our predictions deviate from the current data in case of the branching ratios $\mathcal{B}(D_s^+\rightarrow K^+\eta^\prime)$ and $\mathcal{B}(D^+\rightarrow K^+\eta')$. Future more precise measurements of these channels are therefore highly important in order to clarify the quality of the SU(3)$_F$ expansion in nonleptonic $D\rightarrow P\eta^\prime$ decays.
}

\maketitle
\flushbottom

\section{Introduction \label{sec:introduction}}

Charm decays are a laboratory for the 
exploration of flavor violation in the up quark sector, complementary to the kaon and $b$ physics program. The first evidence of 
non-zero CP asymmetries in $D\to \pi^+\pi^-$ decays \cite{LHCb:2019hro,LHCb:2022lry} received a lot of attention and pushes the focus toward measurements of CP violation. Recent developments for further searches for CP violation are very promising~\cite{LHCb:2023qne,LHCb:2023mwc, LHCb:2023rae, LHCb:2024rkp,CMS:2024hsv, Belle-II:2023vra, Belle:2023bzn}, 
and future prospects are 
bright~\cite{Belle-II:2018jsg, Cerri:2018ypt, LHCb:2018roe}. 
In particular, there has also been recent progress in nonleptonic decays to $\eta^{(\prime)}$ states~\cite{BESIII:2022xhe, Belle:2021dfa, LHCb:2022pxf, LHCb:2021rou, Belle:2021ygw}. 

Predicting hadronic $D$ meson decays is notoriously challenging.
The most prospective method to study hadronic $D$ meson decay amplitudes employs the approximate SU(3)$_F$ symmetry of QCD~\cite{Kingsley:1975fe, Voloshin:1975yx, Barger:1979fu, Golden:1989qx, Savage:1989qr, Buccella:1994nf, Pirtskhalava:2011va, Bhattacharya:2012ah, Hiller:2012xm,  Grossman:2012eb, Cheng:2012wr, Feldmann:2012js, Grossman:2012ry,  Brod:2012ud, Nierste:2015zra, Muller:2015lua, Muller:2015rna, Nierste:2017cua, He:2018php, Grossman:2018ptn, Grossman:2019xcj, Cheng:2019ggx, Dery:2021mll, Bhattacharya:2021ndt, Schacht:2022kuj, Gavrilova:2022hbx, Gavrilova:2023fzy, Gavrilova:2024npn, Iguro:2024uuw} which relates hadronic amplitudes of different decays to each 
other. Control of SU(3)$_F$ breaking effects is important in order to obtain sensitivity to physics beyond the Standard Model (BSM)~\cite{Grossman:2006jg, Dery:2019ysp, Bause:2022jes, Iguro:2024uuw}.
Data on CP asymmetries have no impact on the predictions of branching fractions, because 
CP asymmetries involve different, highly suppressed CKM elements multiplying hadronic matrix elements which do not enter the branching ratios. The opposite is not true, instead a thorough global analysis of 
$D$ branching fractions is needed for the prediction of the CP asymmetries. The plethora of data on branching fractions permits the inclusion of SU(3)$_F$ breaking into such an analysis~\cite{Muller:2015lua} and allows predictions for CP asymmetries which partially include SU(3)$_F$ breaking~\cite{Muller:2015rna}. In Ref.~\cite{Muller:2015lua} $D$ decays into two pseudoscalars were studied, but final states with $\eta$ mesons were not included. In this paper, we focus on $D\to P\eta'$ decays, with $P= \pi, K$ and $\eta$.

In and beyond $D$ meson physics, there are several SU(3)$_F$ studies accommodating $\eta$-$\eta^\prime$ mixing with a universal 
mixing angle $\theta$ describing the rotation of the SU(3)$_F$ eigenstates 
$\eta_0$, $\eta_8$ into the mass eigenstates $\eta$, $\eta^\prime$. However, it is known for a long time that this treatment is inconsistent and gives a poor description of data \cite{Leutwyler:1997yr,Feldmann:1998vh,Bickert:2016fgy, Feldmann:1998sh}. 

This problem has been addressed for the case of decay constants, where the authors in \cite{Feldmann:1998vh} introduce two decay constants and two mixing angles for the description of two matrix elements. For recent calculations with analytical methods and lattice QCD see Refs.~\cite{Gan:2020aco} and \cite{Bali:2021qem}, respectively. As we will show below, this approach needs to be adjusted in order to perform a SU(3)$_F$ study of 
$D\to P\eta$ and $D\to P\eta'$ decays.

In previous analyses of non-leptonic $D$ decays a single, universal mixing angle was used \cite{Bhattacharya:2009ps, Grossman:2012ry, Bhattacharya:2021ndt}.  
In the context of $D$ decays the shortcomings of the single-mixing angle description can be understood as follows: The matrix element $\langle P \eta_8 | H | D \rangle$ of the weak Hamiltonian 
$H$ with some light pseudoscalar meson $P$ is well-defined and related to other $D \to  P P^\prime $ matrix elements by SU(3)$_F$ symmetry. With 
\begin{align}
\ket{\eta_8} &=\ket{\eta}\cos\theta +\ket{\eta^\prime}\sin\theta \label{eq:defth}
\end{align}
one usually employs
\begin{align}
\langle P \eta_8 | H | D \rangle &= 
\cos\theta \langle P \eta | H | D \rangle + \sin\theta \langle P \eta^\prime | H | D \rangle, \label{eq:matrix-element}
\end{align}
in global SU(3)$_F$ analyses of heavy-hadron decays, 
but the description in Eq.~(\ref{eq:matrix-element}) cannot be rigorous, because the masses of $\eta$ and $\eta^\prime$ are very different. These matrix elements are three-point functions  which depend on  three kinematic invariants, namely the squared masses $M_D^2$, $M_P^2$, and $M_{\eta^{(\prime)}}^2$ and thus
will differ due to $M_\eta\neq M_{\eta^\prime}$. This is different from the mass splittings within 
the SU(3)$_F$  octet such as $M_K\neq M_\pi$ which are  SU(3)$_F$-breaking effects and well accommodated by the hadronic parameters describing SU(3)$_F$ breaking in our set-up. On the other hand,
$M_\eta\neq M_{\eta^\prime}$ is unrelated to SU(3)$_F$ breaking and an $\mathcal{O}(1)$ effect 
in the power counting of SU(3)$_F$ breaking. 

We further illustrate this feature with an analogy 
from perturbation theory: One may try to calculate QCD corrections to the  $W^+$-$u_j$-$\bar d_k $ vertex in the basis 
of quark flavour eigenstates and then rotate the quarks into mass eigenstates; the CKM matrix is the analogue of the mixing angle $\theta$ here. This procedure gives the correct result 
as long as quark masses are neglected (i.e. in the limit of exact flavour symmetry),  
while the  correct result depends 
on the masses of the involved quarks. Thus the calculation in the symmetry limit supplemented by 
CKM rotations does not capture the dependence of the vertex function on the masses and momenta of the loop function. 

The purpose of this paper is two-fold: We first explain how $D$ decays into final states with 
$\eta$ or $\eta^\prime$ are treated correctly in SU(3)$_F$ analyses. Then we perform a global analysis of $D\to P \eta^\prime $ branching ratios, with $D=D^0,D^+,D_s^+$, including linear 
SU(3)$_F$ breaking 
to test the quality of SU(3)$_F$ symmetry in these decays and to make predictions for future measurements. 

After discussing $\eta_0$--$\eta_8$ mixing in the context of $D$ decays in Sec.~\ref{sec:eta-mixing}, we give the topological amplitude decomposition for $D\rightarrow P\eta'$ decays in Sec.~\ref{sec:SU3-decomposition}. We present our numerical results of a global fit to current data
in Sec.~\ref{sec:numerics}, before we conclude in Sec.~\ref{sec:conclusions}. Several details as well as the topological diagrams are given in the Appendix.

\section{\boldmath Description of $D$ meson decay matrix elements involving $\eta_{0,8}$ \label{sec:eta-mixing}}
  
We define the octet $\ket{\eta_8}$ and singlet $\ket{\eta_0}$   
states in terms of the corresponding quark bilinears as
\begin{align}
\ket{\eta_8} &= \frac{\ket{u\bar u}+ \ket{d\bar d} -2 \ket{s\bar s}}{\sqrt6}\,, \\
\ket{\eta_0} &= \frac{\ket{u\bar{u}}+\ket{d\bar{d}}+\ket{s \bar{s}}}{\sqrt{3}}\,,
\end{align}
respectively. 
Here, the equality sign should be understood such that the states on the left-hand side and right-hand side share the same quantum numbers.

In the SU(3)$_F$ limit the mass matrix for $(\eta_8,\eta_0)^T$ is diagonal.
SU(3)$_F$ breaking in the strong interaction leads to $\eta_0$--$\eta_8$ mixing,
with physical states $\eta$ and $\eta^{\prime}$. 
After removing unphysical phases, we can write $\ket{\eta_8}$ as 
a real linear combination of $\ket{\eta}$ and $\ket{\eta^{\prime}}$
as in \eq{eq:defth}. Since states of different physical particles are 
orthogonal to each other, $\langle \eta\ket{\eta^{\prime}}=0$, 
one necessarily has  
\begin{align}
\ket{\eta_0} &=-\ket{\eta}\sin\theta +\ket{\eta^\prime}\cos\theta .\label{eq:defe0}
\end{align}
$\theta$ is an SU(3)$_F$-breaking parameter related solely to the spectrum of the QCD hamiltonian and is therefore not directly related to
any observable, because the production mechanism and decay rate of
$\eta$ or $\eta^\prime$ suffers from SU(3) breaking as well.  In the
theoretical prediction of any  observable $\theta$ appears together with
other SU(3)-breaking parameters. If one tries to define the
$\eta$-$\eta^\prime$ mixing angle through measurable quantities, one encounters the
situation that more than one mixing angle
\cite{Leutwyler:1997yr,Feldmann:1998vh} is needed for the theoretical description
and, of course, the such defined angle cannot be immediately used in
other observables.

As mentioned in the introduction, the pitfall of using $\theta$ in
SU(3)$_F$ analyses are the different masses of $\eta$ and $\eta^\prime$.
We may use \eqsand{eq:defth}{eq:defe0} to express
$\ket{P(p_P)\eta (p_\eta)}$  in terms of
$\ket{P(p_P) \eta_{0,8}(p_\eta)}$ and do the same with 
$\ket{P(p_P)\eta^\prime (p_{\eta^\prime}) }$, but the Fock states
$\ket{P(p_P) \eta_{0,8}(p_\eta)}$ and $\ket{P(p_P)
  \eta_{0,8}(p_{\eta^\prime})}$ are different and so are the
corresponding $D$ decay
matrix elements. We write
\begin{align}
  \bra{P\eta}H\ket{D}
  &=\, \cos\theta \bra{P\eta_8} H \ket{D}  -
        \sin\theta \bra{P\eta_0} H \ket{D}     \label{eq:eta-mixing-1}\\
  \bra{P\eta'}H\ket{D}
  &=\,  \sin\theta\bra{P\eta_8} H\ket{D}'+
    \cos\theta \bra{P\eta_0} H\ket{D}'\,, \label{eq:eta-mixing-2}
\end{align}
where 
\begin{align}
\bra{P\eta_8} H\ket{D}' &\neq \bra{P\eta_8} H \ket{D}\,, \label{eq:different-2}\\
\bra{P\eta_0} H \ket{D}' &\neq \bra{P\eta_0} H \ket{D}\,. \label{eq:different-3}
\end{align}
We emphasize that Eqs.~(\ref{eq:eta-mixing-1}) and (\ref{eq:eta-mixing-2}) clarify and correct Eq.~(\ref{eq:matrix-element}).

In particular, the departure of
\eqsand{eq:different-2}{eq:different-3} from equalities is
an $\mathcal{O}(1)$ effect,~\emph{i.e.},~not suppressed by SU(3)$_F$
  breaking. Numerically, one finds $m_{\eta^\prime} = 0.95778$ GeV and
  $m_{\eta} = 0.547862$ GeV, so that the kinematical invariants
  $p_{\eta^{(\prime)}}^2=m_{\eta^{(\prime)}}^2$ entering
  \eqsand{eq:different-2}{eq:different-3}
  differ by more than a factor of 3.

Note that this fact leads to $D\rightarrow P\eta'$ decays being uncorrelated to 
$D\rightarrow P\eta$ decays in SU(3)$_F$ analyses, because the
  corresponding matrix elements are unrelated under SU(3)$_F$.  This
could only be changed by putting model-dependent assumptions about the scaling of the matrix elements w.r.t.\ kinematical variables
into place.  Here, we refrain from making such assumptions and therefore
focus on $D\rightarrow P\eta'$ decays only. A study of
$D\rightarrow P\eta$ decays in conjunction with all $D\rightarrow PP^\prime$ modes is left for
future work.

For the mixing between the states, we assume that the mixing angle
  is linear in the  SU(3)$_F$ breaking parameter $\varepsilon\sim 30\%$,
entailing
\begin{align}
\theta &= \mathcal{O}(\varepsilon)\,, \label{eq:SU3-limit-1}
\end{align}
and implying
\begin{align}
\sin\theta &= \mathcal{O}(\varepsilon)\,,\\
\cos\theta &= 1 - \mathcal{O}(\varepsilon^2)\,, \label{eq:cos-expansion}
\end{align}
and
\begin{align}
\ket{\eta}  &= \ket{\eta_8} + \mathcal{O}(\varepsilon)\,, \label{eq:SU3-limit-2}\\
\ket{\eta'} &= \ket{\eta_0} + \mathcal{O}(\varepsilon)\,. \label{eq:SU3-limit-3}
\end{align}
As $\sin\theta$ always appears together with the $\mathcal{O}(1)$ hadronic matrix elements $\bra{P\eta_8} H \ket{D}^\prime$,
we will absorb the mixing angles into the matrix elements in our fit. 

Our approach to consider the $D\to\eta' P$ decays separately from the $D\to P\eta$ and $D\to PP$ decays within an $SU(3)_F$ analysis completely differs from previous studies of non-leptonic $D$ decays. We stress that our approach is also genuinely different from the treatment of $\eta^{(\prime)}$-decay constants presented in Ref.~\cite{Feldmann:1998vh}. In Ref.~\cite{Feldmann:1998vh} different mixing angles $\theta_8$ and $\theta_0$ are introduced for matrix elements of octet and singlet currents. While in these matrix elements the current operator is octet or singlet, our matrix elements of the weak hamiltonian $H$ in \eqsand{eq:eta-mixing-1}{eq:eta-mixing-2} involve different representations of  $SU(3)_F$. If one wanted to adapt the approach of  Ref.~\cite{Feldmann:1998vh} to our case, one must first decompose these matrix 
elements into irreducible representations via the Wigner-Eckhart theorem and then define a different mixing angle for each reduced matrix element. Furthermore, the approach of 
Ref.~\cite{Feldmann:1998vh} does not introduce matrix elements of $\eta_0$ or 
$\eta_8$ which are needed to relate  $\bra{P\eta_8}H\ket{D}$ to the other $\bra{P P^\prime}H\ket{D}$ matrix elements. As another complication,  
compared to decay constants, the hadronic matrix elements in Eqs.~(\ref{eq:eta-mixing-1}) and (\ref{eq:eta-mixing-2}) depend on the flavor of the decaying meson ($D_s^+$, $D^0$, or $D^+$) and on the flavour of the pseudoscalar meson $P$, which is
an $SU(3)_F$ breaking effect. Accommodating this effect by choosing different mixing angles for $D_s^+$, $D^0$, and $D^+$ matrix elements further impedes a global SU(3)$_F$ analysis.
Our approach permits a bookkeeping of all $SU(3)_F$ breaking effects
stemming from the decay matrix elements or $\eta-\eta^\prime$ mixing as needed for 
a global analysis of decays with and without $\eta$ in the final state. The price to pay is an $\eta-\eta^\prime$ mixing angle which is not directly related to an observable. In our analysis $\sin\theta$ always appears in combination with  matrix elements $\bra{P\eta_8} H \ket{D}^\prime$ and only their product can be determined.

\section{SU(3)$_F$ Decomposition of $D\to P \eta'$ decays \label{sec:SU3-decomposition}}

\subsection{Topological Decomposition}
\begin{figure}[t!]
\begin{center}
\subfigure[\, $T_{18}^{}$]{
        \includegraphics[width=0.2\textwidth]{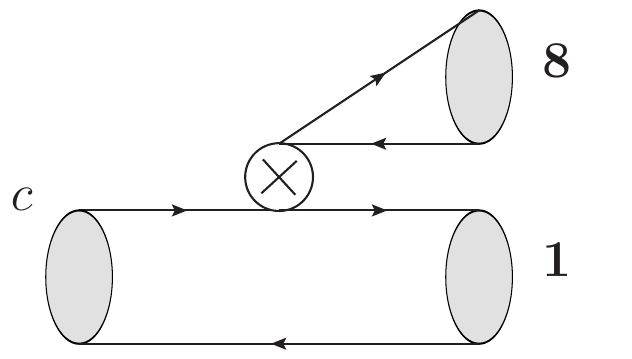}
}
\hfill %
\subfigure[\, $A_{18}$]{
        \includegraphics[width=0.2\textwidth]{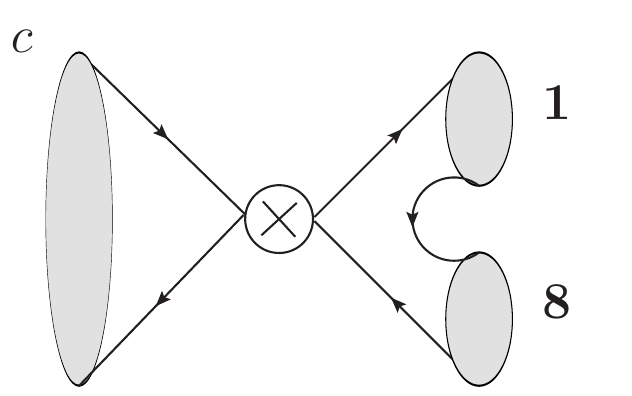}
}
\hfill %
\subfigure[\, $A_{81}$]{
        \includegraphics[width=0.2\textwidth]{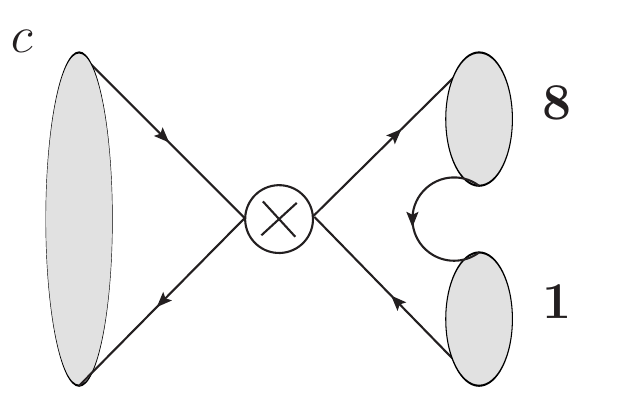}
}
\hfill %
\subfigure[\, $A_{H}$]{
        \includegraphics[width=0.2\textwidth]{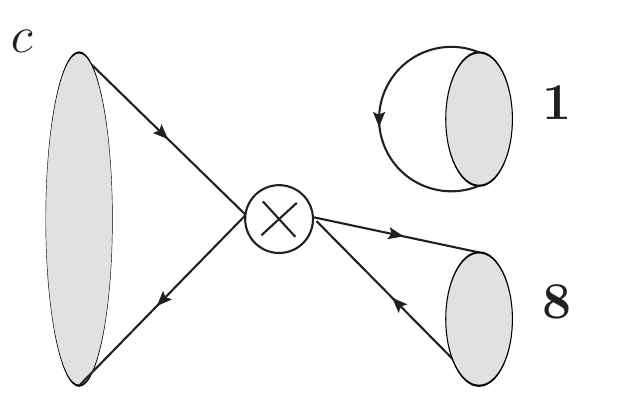}
}
\\
\subfigure[\, $C_{81}^{}$]{
        \includegraphics[width=0.2\textwidth]{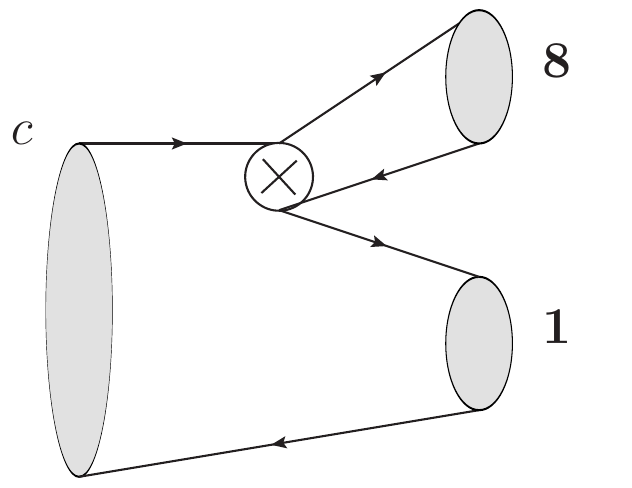}
}
\hfill %
\subfigure[\, $E_{18}$]{
        \includegraphics[width=0.2\textwidth]{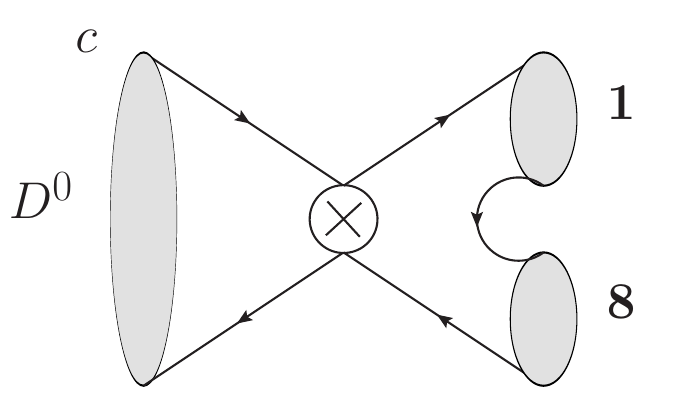}
}
\hfill %
\subfigure[\, $E_{81}^{}$]{
        \includegraphics[width=0.2\textwidth]{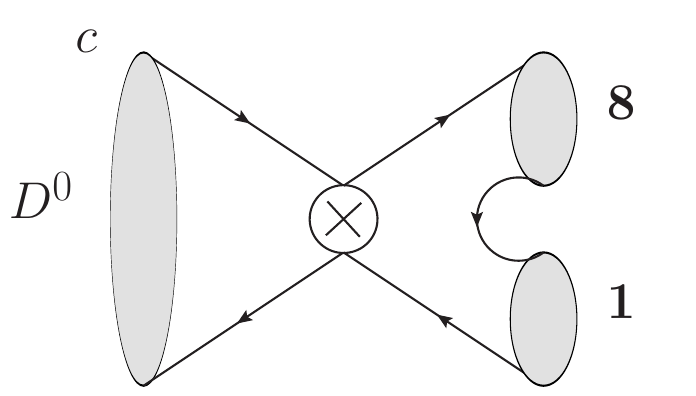}
}
\hfill %
\subfigure[\, $E_{H}$]{
        \includegraphics[width=0.2\textwidth]{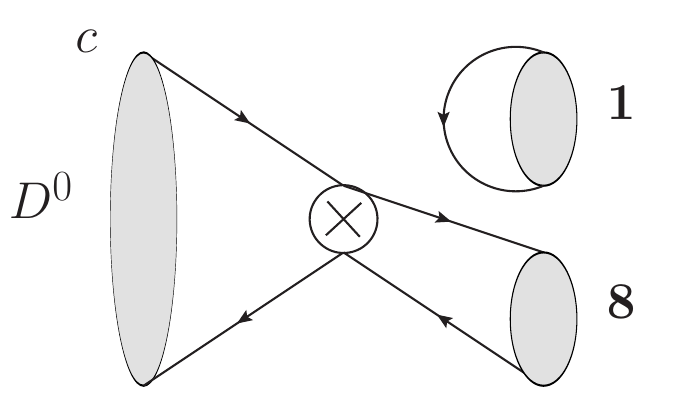}
}
\end{center}
\caption{
Topologies that contribute to $D\to P \eta'$ in the SU(3)$_F$ limit. We distinguish whether the $\eta_1$ singlet is formed from the outgoing quark or antiquark with the labels $18$ and $81$, respectively. The topology $T_{81}$ does not exist due to charge conservation. The diagram $C_{18}$ does exist but does not enter here because the weak interaction produces the quark-antiquark pair in a $U$-spin triplet, while the $\eta_0$ is a singlet state. The index $H$ denotes the hairpin diagrams. We note that hairpin diagrams also accommodate possible glueball components of the $\eta_0$.
 \label{fig:su3lim}}
\end{figure}

Our conventions for the meson states are given in Appendix~\ref{app:conv}. In the \SUt{} limit and neglecting CKM-subleading contributions, the $D\to P\eta'$ decays where $P=\pi, K, \eta$ are parametrized by the topological tree ($T$), annihilation ($A$), colour-suppressed ($C$) and exchange ($E$) like amplitudes shown in Fig.~\ref{fig:su3lim}. The circle with a cross denotes the $W$-boson exchange of the weak interaction. In addition, in the \SUt{} analysis, when breaking effects are included, it is important to differentiate whether the $\eta_1$ singlet is formed from the outgoing quark or antiquark, which we label with indices 18 and 81, respectively. We note that this distinction is rarely seen in the literature, but also only necessary when including SU(3)$_F$ breaking effects. This will become more clear through the redefinitions in Eqs.~(\ref{eq:redefsu3}) and (\ref{eq:redefsu3-2}) below. We also note that $T_{81}$ does not exist due to charge conservation. The topology $C_{18}$ does exist, but does not enter due to the CKM structure of the decays. For $A$ and $E$, we need also to include the corresponding hairpin diagrams, which we label by the index $H$. 

We write the amplitudes of the Cabibbo-favoured (CF), singly Cabibbo-suppressed (SCS) and doubly Cabibbo-suppressed (DCS) $D\to P \eta'$ decays as~\cite{Muller:2015lua}
\begin{align}
\label{eq:ampli}
    \mathcal{A}^\text{CF}(d) &\equiv V_{cs}^*V_{ud}\mathcal{A}(d)\equiv V_{cs}^*V_{ud}\sum_ic_i^d \mathcal{T}_i ,\nonumber\\
    \mathcal{A}^\text{SCS}(d) &\equiv \lambda_{sd}\mathcal{A}(d)\equiv \lambda_{sd}\sum_ic_i^d \mathcal{T}_i, \nonumber\\
    \mathcal{A}^\text{DCS}(d) &\equiv V_{cd}^*V_{us}\mathcal{A~}(d)\equiv V_{cd}^*V_{us}\sum_ic_i^d \mathcal{T}_i, 
\end{align}
where $\lambda_{sd} = (\lambda_s - \lambda_d)/2 = (V_{cs}^*V_{us} - V_{cd}^*V_{ud})/2 \simeq \lambda_s\simeq -\lambda_d$. In this work, we neglect CKM-subleading effects, \emph{i.e.} we set $\lambda_b = V_{cb}^*V_{ub} =0$, as they have a negligible effect on the branching ratios. For the different decay modes $d=D\to P \eta'$, the coefficients $c_i^d$ multiply the different topological amplitudes $\mathcal{T}_i$. In the \SUt-limit, these coefficients are listed in Table~\ref{tab:topoparametrizationsu3lim}.

\begin{table}[t!]
\begin{center}
\begin{tabular} 
{c|cccccccc}
\hline \hline
Decay ampl. {$\mathcal{A}(d)$} & 
$T_{18}$ &   
$A_{18}$ &  $A_{81}$ & $A_{H}$ &
 $C_{81}$ & 
$E_{18}$ & $E_{81}$ & $E_{H}$ \\\hline\hline
\multicolumn{9}{c}{SCS} \\\hline\hline
$\mathcal{A}( D^0 \rightarrow \pi^0 \eta' )$  &   
$0$ &   
$0$ &  $0$  & $0$ &
 $\frac{1}{\sqrt{6}}$ & 
$\frac{1}{\sqrt{6}}$ & $\frac{1}{\sqrt{6}}$  & $\sqrt{\frac{3}{2}}$ \\\hline
$\mathcal{A}( D^0 \rightarrow \eta \eta' )$  & 
$0$ &   
$0$ &  $0$  & $0$ &
 $-\frac{1}{\sqrt{2}}$ & 
$-\frac{1}{\sqrt{2}}$ & $-\frac{1}{\sqrt{2}}$ &  $-\frac{3}{\sqrt{2}}$\\\hline
$\mathcal{A}( D^+ \rightarrow \pi^+ \eta' )$ & 
$-\frac{1}{\sqrt{3}}$ &   
$-\frac{1}{\sqrt{3}}$ &  $-\frac{1}{\sqrt{3}}$  & $-\sqrt{3}$ &
 $0$ & 
$0$ & $0$ &  $0$\\\hline
$\mathcal{A}( D_s^+ \rightarrow K^+ \eta' )$ & 
$\frac{1}{\sqrt{3}}$ &   
$\sqrt{\frac{1}{3}}$ &  $\frac{1}{\sqrt{3}}$ & $\sqrt{3}$ &
 $0$ & 
$0$ & $0$ &  $0$\\\hline\hline

\multicolumn{9}{c}{CF} \\\hline\hline
$\mathcal{A}( D^0 \rightarrow \overline{K}^0 \eta')$ & 
$0$ &   
$0$ &  $0$  & $0$ &
 $\frac{1}{\sqrt{3}}$ & 
$\frac{1}{\sqrt{3}}$ & $\frac{1}{\sqrt{3}}$ & $\sqrt{3}$
\\\hline

$\mathcal{A}( D_s^+ \rightarrow \pi^+ \eta' )$  & 
$\frac{1}{\sqrt{3}}$ &   
$\frac{1}{\sqrt{3}}$ &  $\frac{1}{\sqrt{3}}$  & $\sqrt{3}$ &
 $0$ & 
$0$ & $0$ & $0$\\\hline\hline

\multicolumn{9}{c}{DCS} \\\hline\hline

$\mathcal{A}( D^0 \rightarrow K^0 \eta' )$ & 
$0$ &   
$0$ &  $0$  & $0$ &
 $\frac{1}{\sqrt{3}}$ & 
$\frac{1}{\sqrt{3}}$ & $\frac{1}{\sqrt{3}}$  & $\sqrt{3}$\\\hline

$\mathcal{A}( D^+ \rightarrow K^+ \eta' )$  &  
$\frac{1}{\sqrt{3}}$ &   
$\frac{1}{\sqrt{3}}$ &  $\frac{1}{\sqrt{3}}$ & $\sqrt{3}$ &
 $0$ & 
$0$ & $0$ & $0$\\\hline\hline

\end{tabular}
\caption{SU(3)$_F$ limit decomposition of $D\rightarrow P\eta'$ decays.} 
\label{tab:topoparametrizationsu3lim}
\end{center}
\end{table}

Employing the same normalization as in Ref.~\cite{Muller:2015lua}, the branching ratio is then obtained as
\begin{align}\label{eq:br}
    \mathcal{B}(D\to P \eta') &= |\mathcal{A}^X(D\to P \eta')|^2 \times \mathcal{P}(D\to P \eta')\,,
    \end{align}
with the phase-space function
    \begin{align}
    \mathcal{P}(D\to P \eta') &=  \frac{ \tau_D }{16\pi m_D^3} \times \sqrt{(m_D^2 - (m_P - m_{\eta'})^2)(m_D^2 - (m_P + m_{\eta'})^2)}\,,
\end{align}
and where $X$ refers to CF, SCS and DCS, indicating the CKM suppression given in Eq.~\eqref{eq:ampli}. 

From Table~\ref{tab:topoparametrizationsu3lim}, we observe that in the \SUt-limit the eight $D\to P\eta'$ decays are split into two sections. The table has rank two, meaning that there are linear dependences in our parameterization. Absorbing these linear dependent parameters, we can redefine the ``tree'' and ``color-suppressed'' parameters: 
\begin{align}\label{eq:redefsu3}
    \hat{T}_{18} = & T_{18} + A_{18} + A_{81} + 3 A_H\,, \\
    \hat{C}_{81} = & C_{81} + E_{18} + E_{81} + 3 E_H\,. \label{eq:redefsu3-2}
\end{align}
These linear dependences are also reflected in two sets of amplitude level sum rules, namely 
\begin{align}
    \mathcal{A}(D^+\to K^+ \eta') &= \mathcal{A}(D_s^+\to \pi^+ \eta') 
    = \mathcal{A}(D_s^+\to K^+ \eta') = - \mathcal{A}(D^+\to \pi^+ \eta')\,, 
\end{align}
and
\begin{align}
    \mathcal{A}(D^0\to K^0 \eta') &= \mathcal{A}(D^0\to \overline{K}^0 \eta') 
   = \sqrt{2} \mathcal{A}(D^0\to \pi^0 \eta') = -\sqrt{\frac{2}{3}}\mathcal{A}(D^0\to \eta \eta')\,.
\end{align}
The experimental branching ratios allow for a direct test of the \SUt-limit sum rules when correcting for phase-space effects and CKM factors, which we show below in Sec.~\ref{sec:2d}. In principle, the $T_{18}$ amplitude could be estimated in the large $N_c$ limit as done in Ref.~\cite{Muller:2015rna}. However, $T_{18}$ cannot be extracted unambiguously by itself from experimental data due to the redefinition in Eq.~\eqref{eq:redefsu3} and therefore such a comparison with theoretical estimates is not feasible.

\subsection{Linear \SUt-breaking}

We include linear \SUt{} breaking following the formalism of Refs.~\cite{Gronau:1995hm, Muller:2015lua}. The \SUt-breaking part of the Hamiltonian is given as
\begin{equation}\label{eq:Hsut}
    H_{\cancel{\mathrm{SU(3)_F}} }= (m_s-m_d) \bar{s}{s} \ ,
\end{equation}
for which the Feynman rule is denoted by a cross on the $s$-quark line. Furthermore, we denote the linear \SUt-breaking topologies with a superscript  \lq\lq{}$(1)$\rq\rq{}. The diagrammatic definitions of the \SUt-breaking topologies are given in Appendix~\ref{app:su3break}. The perturbation $ H_{\cancel{\mathrm{SU(3)_F}}}$ also introduces the $\eta$--$\eta'$ mixing. Therefore, we treat the $\eta_8$ contribution to the $\eta'$ at the same level in the power counting. Such contributions, coming from the octet contribution to the $\eta'$, are labelled with the subscript \lq\lq{}88\rq\rq{}.

At the first order of the expansion in SU(3)$_F$-breaking effects also appears the broken penguin~\cite{Brod:2012ud},~\emph{i.e.}~the combination of penguin-contractions of the tree operator $P_\text{break}\equiv P_s-P_d$. Note that penguin annihilation diagrams do not contribute at the order we consider here.

We give the corresponding SU(3)$_F$-breaking decomposition of the eight $D\to P\eta'$ decays in Appendix~\ref{app:su3break}. As in the \SUt-limit case, there are linear dependent columns, meaning that our parametrization contains redundant parameters which can not be 
disentangled through a fit to the data. The combination of the SU(3)$_F$ limit and first order SU(3)$_F$ breaking matrix has rank six, smaller than the number of 
parameters. 

As it is not possible to determine these parameters from theory calculations, we redefine several parameters in order to remove flat directions in the fit as much as possible. We identify the flat directions by calculating the nullspace of the SU(3)$_F$-breaking matrix, see Refs.~\cite{Hiller:2012xm, Muller:2015lua} for more details.

\begin{table}[t]
    \centering
    \begin{tabular}{c|ccccccccc}\hline\hline
 Decay ampl. {$\mathcal{A}(d)$}  & $\hat{T}_{18}$ & $\hat{T}_{18,1}^{(1)}$ & $\hat{T}_{18,2}^{(1)}$ & $\hat{T}_{88}^{(1)}$ & $\hat{C}_{18,1}^{(1)}$ & $\hat{C}_{81}$ & $\hat{C}_{81,1}^{(1)}$ & $\hat{C}_{81,2}^{(1)}$ & $\hat{C}_{88}^{(1)}$\\ \hline\hline
\multicolumn{10}{c}{SCS} \\\hline\hline
 $\mathcal{A}( D^0 \rightarrow \pi^0 \eta' )$ & $0$ & $0$ & $0$ & $0$ & $\frac{1}{\sqrt{6}}$ & $\frac{1}{\sqrt{6}}$ & $0$ & $0$ & $-\frac{1}{\sqrt{3}}$  \\
 $\mathcal{A}( D^0 \rightarrow \eta \eta' )$  & $0$ & $0$ & $0$ & $0$ & $\frac{1}{3 \sqrt{2}}$ & $-\frac{1}{\sqrt{2}}$ & $-\frac{\sqrt{2}}{3}$ & $-\frac{\sqrt{2}}{3}$ & $-1$ \\
$\mathcal{A}( D^+ \rightarrow \pi^+ \eta' )$ &  $-\frac{1}{\sqrt{3}}$ & $0$ & $0$ & $-\frac{1}{\sqrt{6}}$ & $\frac{1}{\sqrt{3}}$ & $0$ & $0$ & $0$ & $-\sqrt{\frac{3}{2}}$  \\
 $\mathcal{A}( D_s^+ \rightarrow K^+ \eta' )$ &  $\frac{1}{\sqrt{3}}$ & $\frac{1}{\sqrt{3}}$ & $\frac{1}{\sqrt{3}}$ & $-\sqrt{\frac{2}{3}}$ &$\frac{1}{\sqrt{3}}$ & $0$ & $0$ & $0$ & $-\sqrt{\frac{3}{2}}$   \\\hline\hline
\multicolumn{10}{c}{CF} \\\hline\hline
$\mathcal{A}( D^0 \rightarrow \overline{K}^0 \eta')$ &  $0$ & $0$ & $0$ & $0$ & $0$ & $\frac{1}{\sqrt{3}}$ & $\frac{1}{\sqrt{3}}$ & $0$ & $\frac{1}{\sqrt{6}}$ \\
$\mathcal{A}( D_s^+ \rightarrow \pi^+ \eta' )$  & $\frac{1}{\sqrt{3}}$ & $\frac{1}{\sqrt{3}}$ & $0$ & $-\sqrt{\frac{2}{3}}$ & $0$ & $0$ & $0$ & $0$ & $0$\\\hline\hline
\multicolumn{10}{c}{DCS} \\\hline\hline
$\mathcal{A}( D^0 \rightarrow K^0 \eta' )$ & $0$ & $0$ & $0$ & $0$ & $0$ & $\frac{1}{\sqrt{3}}$ & $0$ & $\frac{1}{\sqrt{3}}$ & $\frac{1}{\sqrt{6}}$ \\
$\mathcal{A}( D^+ \rightarrow K^+ \eta' )$  &  $\frac{1}{\sqrt{3}}$ & $0$ & $\frac{1}{\sqrt{3}}$ & $\frac{1}{\sqrt{6}}$ & $0$ & $0$ & $0$ & $0$ & $0$ \\\hline\hline
    \end{tabular}
    \caption{SU(3)$_F$-breaking decomposition of $D\rightarrow P\eta'$ decays including parameter redefinitions.}
    \label{tab:hatparam}
\end{table}

We find the following redefinitions:
\begin{align}
    \hat{T}_{18,1}^{(1)} = & T_{18,1}^{(1)} + T_{18,3}^{(1)} + A_{18,1}^{(1)} + A_{81,1}^{(1)} + 3 A_{H,1}^{(1)} + 3\sqrt{2} A_{88}^{(1)} + 3\sqrt{2} E_{88}^{(1)}\,, \label{eq:redefinitions-first}\\
    \hat{T}_{18,2}^{(1)} = & T_{18,2}^{(1)} + A_{18,2}^{(1)} + A_{81,2}^{(1)} + A_{81,3}^{(1)} + 3 A_{H,2}^{(1)} -\frac{3}{\sqrt{2}} A_{88}^{(1)} -\frac{3}{\sqrt{2}} E_{88}^{(1)}\,, \\
    \hat{T}_{88}^{(1)} = & T_{88}^{(1)}+2A_{88}^{(1)}+3E_{88}^{(1)}\,,\\
    \hat{C}_{18,1}^{(1)} = & C_{18,1}^{(1)} + C_{18,2}^{(1)} + P_{18,\mathrm{break}}^{(1)} + P_{81,\mathrm{break}}^{(1)}\,, \\
    \hat{C}_{81,1}^{(1)} = & C_{81,1}^{(1)} + E_{18,1}^{(1)} + E_{18,3}^{(1)} + E_{81,1}^{(1)} + 3 E_{H,1}^{(1)}\,, \\
    \hat{C}_{81,2}^{(1)} = & C_{81,2}^{(1)} + E_{18,2}^{(1)} + E_{81,2}^{(1)} + E_{81,3}^{(1)} + 3 E_{H,2}^{(1)}\,, \\
    \hat{C}_{88}^{(1)} = & C_{88}^{(1)}-E_{88}^{(1)}\,. \label{eq:redefinitions-last}
\end{align}
In terms of these parameters, the matrix in Appendix~\ref{app:su3break} can be equivalently, but much simpler, be written as the matrix in Table~\ref{tab:hatparam}. 

In principle, additional redefinitions would be possible, however these would mix SU(3)$_F$ limit parameters with SU(3)$_F$-breaking matrix elements. In order to keep the power counting simple, we decide not to perform these additional redefinitions.

For these eight decays, we find two sum rules that also hold with \SUt{} breaking: 
\begin{align}\label{eq:sr1}
&    \mathcal{A}(D^+\to K^+ \eta') + \mathcal{A}(D_s^+\to \pi^+ \eta') - \mathcal{A}(D_s^+\to K^+ \eta') + \mathcal{A}(D^+\to \pi^+ \eta') = 0\,,
\end{align}
and
\begin{align}\label{eq:sr2}
    \mathcal{A}(D^0\to K^0 \eta') + \mathcal{A}(D^0\to \overline{K}^0 \eta') - \
 \frac{\mathcal{A}(D^0\to \pi^0 \eta')}{\sqrt{2}} +\sqrt{\frac{3}{2}}\mathcal{A}(D^0\to \eta \eta') = 0\,,
\end{align}
in agreement with Ref.~\cite{Grossman:2012ry}. The number of sum rules together with the number of decay channels determines the number of linearly independent parameters. Therefore, the above sum rules also imply agreement with the matrix rank of the corresponding coefficient tables found in the group-theoretical approach in Ref.~\cite{Grossman:2012ry}.

\section{Numerical Results \label{sec:numerics}}

\subsection{Constraining diagrammatic SU(3)$_F$ breaking}
We impose constraints on the SU(3)$_F$-breaking part of the amplitude through
\begin{align}
\left|\frac{\mathcal{A}_{\cancel{\mathrm{{SU(3)}}}}(D\to P \eta')}{\mathcal{A}_{\mathrm{{SU(3)-lim}}}(D\to P \eta')}\right|\leq \varepsilon\,, \label{eq:SU3-breaking}
\end{align}
where $\mathcal{A}_{\cancel{\mathrm{SU(3)}}}$ corresponds to the part of the amplitude arising only from the \SUt{}-breaking terms in the amplitude, $\mathcal{A}_{\mathrm{SU(3)-lim}}$ refers to the \SUt~limit part of the amplitude, and $\varepsilon$ is the imposed amount of allowed SU(3)$_F$ breaking. This constraint is applied to all eight $D\to P \eta'$ channels. We note that our treatment of \SUt~breaking differs from Refs.~\cite{Hiller:2012xm,Muller:2015rna}, where also the \SUt{}-breaking of individual parameters has been constrained. Due to the redefinitions in Eq.~\eqref{eq:redefinitions-last}, the different topologies mix to a large amount, making such a constraint less intuitive. Therefore, we only constrain the effect of \SUt~breaking on an amplitude as a whole, and not its individual contributions. 
Commonly it is expected that $\varepsilon \sim 20\%$--$30\%$, 
motivated from the ratio of decay constants $f_K/f_{\pi}-1\sim 20\%$~\cite{FlavourLatticeAveragingGroupFLAG:2021npn}.

We emphasize that our constraint on SU(3)$_F$-breaking in Eq.~\eqref{eq:SU3-breaking} requires the separation of SU(3)$_F$-breaking and SU(3)$_F$-limit parameters.
This is possible as we do not mix these two contributions in our redefinitions Eqs.~(\ref{eq:redefsu3}), (\ref{eq:redefsu3-2}), and (\ref{eq:redefinitions-first})--(\ref{eq:redefinitions-last}). 
Importantly, the form of the employed constraint on SU(3)$_F$ breaking and of the redefinitions ensure that our redefinitions are merely a technical way to simplify the fit and do not affect the fit results for observables, \emph{i.e.}~branching ratios.

\subsection{Fit setup and \SUt~test}
\begin{table}[t]
    \centering
    \begin{tabular}{ccc}\hline\hline
        Observable &  Global theory fit      & Experiment  \\\hline
        $\mathcal{B}(D^0\to \pi^0\eta' )$  & $(9.15^{+1.00}_{-0.99})\cdot 10^{-4}$ &  $(9.2\pm1.0)\cdot 10^{-4}$  \\
         $\mathcal{B}(D^0\to \eta \eta')$ & $(0.98 \pm 0.18)\cdot 10^{-3}$ &  $(1.01 \pm 0.19)\cdot 10^{-3}$  \\
         $\mathcal{B}(D^+\to \pi^+\eta' )$ & $(4.95 \pm 0.19)\cdot 10^{-3}$&  $(4.97 \pm 0.19)\cdot 10^{-3}$ \\
         $\mathcal{B}(D_s^+\to K^+\eta')$ & $(2.22 \pm 0.17)\cdot 10^{-3}$ &  $(2.64 \pm 0.24)\cdot 10^{-3}$  \\
         $\mathcal{B}(D^0\to K_S \eta')$ & $(9.56 ^{+0.27} _{-0.25})\cdot 10^{-3}$ &  $(9.49 \pm 0.32)\cdot 10^{-3}$   \\
         $\mathcal{B}(D^0\to K_L \eta')$ & $(8.04 ^{+0.26}_{-0.29})\cdot 10^{-3}$ &  $(8.12 \pm 0.35)\cdot 10^{-3}$   \\
         $\mathcal{B}(D_s^+\to \pi^+ \eta')$ & $(4.17 \pm 0.23)\cdot 10^{-2}$  &  $(3.94 \pm 0.25)\cdot 10^{-2}$  \\
         $\mathcal{B}(D^+\to K^+ \eta')$ & $(2.11 \pm0.17)\cdot 10^{-4}$ &   $(1.85 \pm 0.20)\cdot 10^{-4}$  \\\hline\hline
    \end{tabular}
    \caption{Comparison of experimental input data from Ref.~\cite{ParticleDataGroup:2024cfk} to our global theory fit allowing for 30\% SU(3)$_F$ breaking. All shown errors are $1\sigma$ uncertainties. The experimental data are uncorrelated with the exception of $\mathcal{B}(D^+\to K^+ \eta')$ and $\mathcal{B}(D^+\to \pi^+ \eta')$, which have a correlation coefficient of $0.32$ \cite{ParticleDataGroup:2024cfk}.}  \label{tab:inputdata}
\end{table}

\begin{figure}[t]
\centering
  \includegraphics[width=0.6\textwidth]{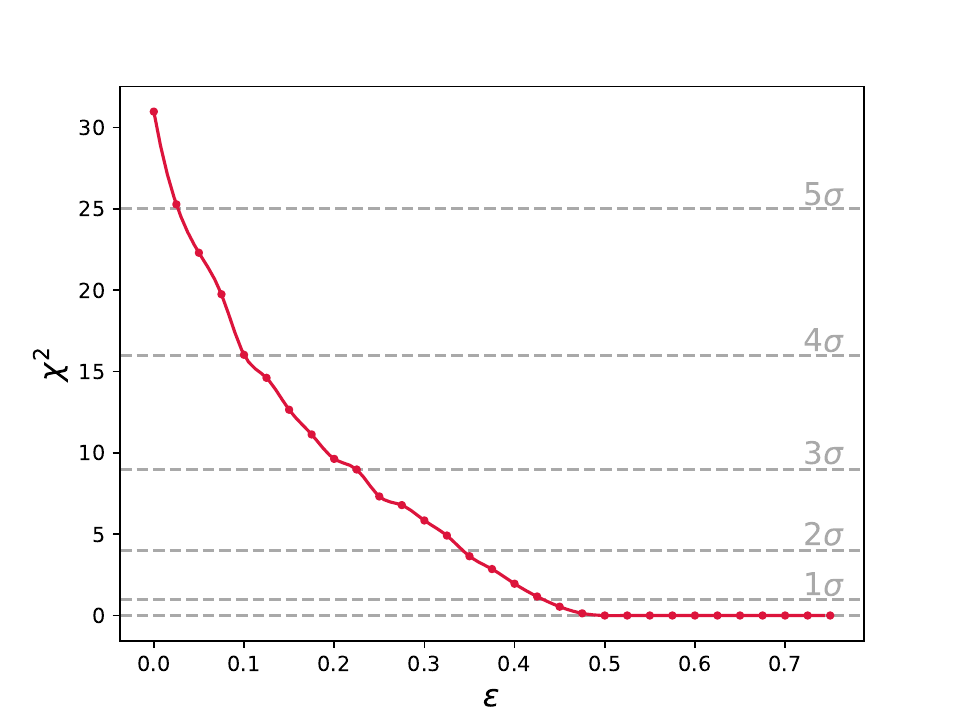}
  \caption{Minimum $\chi^2$ value as a function of the imposed \SUt~breaking constraint $\varepsilon$ at the amplitude level. Starting at $\varepsilon=50\%$, there is a perfect description of the data.}
  \label{fig:su3bscan}
\end{figure}

In our global fit, we use the measured branching ratios given in Table~\ref{tab:inputdata}. Masses and lifetimes are taken from Ref.~\cite{ParticleDataGroup:2024cfk} and for the CKM factors we use the Wolfenstein parametrization up to $\lambda^2$ with $\lambda = 0.225$ \cite{ParticleDataGroup:2024cfk}. The theoretical parametrization for the amplitudes is given in Table~\ref{tab:hatparam}, giving in total 17 fit parameters for eight observables. In the \SUt-limit, we have only three parameters
\begin{align}
    |\hat{T}_{18}|, \quad |\hat{C}_{81}/\hat{T}_{18}|, \quad {\rm arg} (\hat{C}_{81}) \ ,  
\end{align}
where the absolute phase of $\hat{T}_{18}$ is undetermined and without loss of generality set to zero. For convenience, we normalize all parameters to $\hat{T}_{18}$. For the \SUt-breaking parameters, we then have
\begin{align}
   \quad |\hat{T}_{18,1}^{(1)}/\hat{T}_{18}|, \quad {\rm arg} (\hat{T}^{(1)}_{18,1}), \quad |\hat{T}_{18,2}^{(1)}/\hat{T}_{18}|, \quad {\rm arg} (\hat{T}^{(1)}_{18,2}), \quad |\hat{T}_{88}^{(1)}/\hat{T}_{18}|, \quad {\rm arg} (\hat{T}^{(1)}_{88}),
\end{align}
 and 
\begin{align}
   \quad |\hat{C}_{81,1}^{(1)}/\hat{T}_{18}|, \quad {\rm arg} (\hat{C}^{(1)}_{81,1}), \quad |\hat{C}_{81,2}^{(1)}/\hat{T}_{18}|, \quad {\rm arg} (\hat{C}^{(1)}_{81,2}), \nonumber \\
   \quad |\hat{C}_{18,1}^{(1)}/\hat{T}_{18}|, \quad {\rm arg} (\hat{C}^{(1)}_{18,1}), \quad \quad |\hat{C}_{88}^{(1)}/\hat{T}_{18}|, \quad {\rm arg} (\hat{C}^{(1)}_{88}).
\end{align}

We then perform a $\chi^2$ minimisation by constructing
\begin{align}
\chi^2 &= (\vec{y}_\text{data}-\vec{y}_\text{theo})^T \,\text{Cov}^{-1}\,(\vec{y}_\text{data}-\vec{y}_\text{theo})\,. 
\end{align}
Every minimisation is performed 100 times starting from randomised initial starting points for the parameters in order to avoid local minima of the function. We include the \SUt-breaking constraints of Eq.~\eqref{eq:SU3-breaking} in a frequentist analysis using the Sequential Least SQuares Programming (SLSQP) algorithm implemented in SciPy  \cite{kraft1988software,2020SciPy-NMeth}.

To test our \SUt~assumption, we first perform a scan of the value of  $\chi^2$ with respect to the imposed \SUt~-breaking constraint, $\varepsilon$. This scan is shown in Fig.~\ref{fig:su3bscan}, where we also indicate $\chi^2 = 1,4,9,16,25$ lines. We note that the identification of those lines with CLs is ambiguous, as the number of degrees of freedom is not well-defined. In the following we count the constraint on SU(3)$_F$ breaking as one degree of freedom.  This procedure could be improved using toy Monte Carlo data with a Feldman-Cousins approach, which is beyond the scope of our paper. We obtain a perfect fit to the experimental data for $\varepsilon \geq 50\%$ where we obtain $\chi^2=0$. At the same time, we note that such a large amount of \SUt{} breaking in principle violates our initial assumption to consider only linear \SUt{} breaking. The SU(3)$_F$-limit fit, where $\varepsilon=0$, has $\chi^2=31.4$, about half of this comes from the $D^0\to \pi^0\eta'$ decay, which we comment on later. 
Comparing to the scenario that gives $\chi^2=0$ ($\varepsilon \geq 50\%$) excludes the \SUt{} limit therefore at $5.6\sigma$.  
It is therefore clear that SU(3)$_F$-breaking has to be taken into account. 

Our nominal result corresponds to $\varepsilon=30\%$, with $\chi^2=6.01$, in slight tension of $2.5\sigma$ with the data when compared to the fit with $\varepsilon=50\%$ (i.e. $\chi^2=0$).

\begin{figure}[t]
  \begin{minipage}[c]{0.48\textwidth}
    \includegraphics[width=\textwidth]{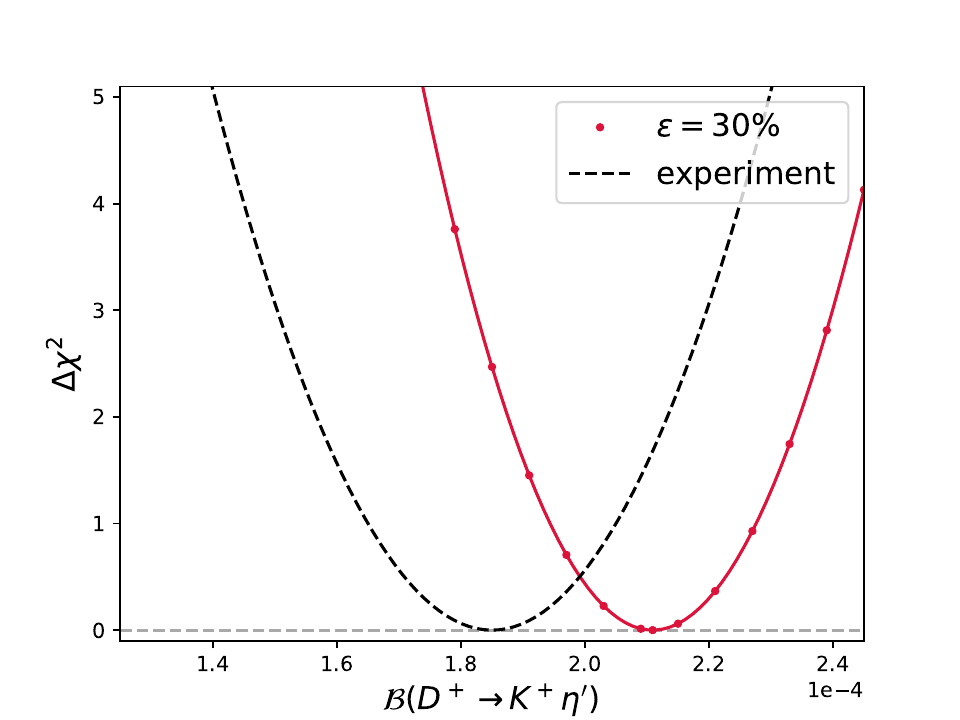}
  \end{minipage}\hfill
  \begin{minipage}[c]{0.48\textwidth}
    \includegraphics[width=\textwidth]{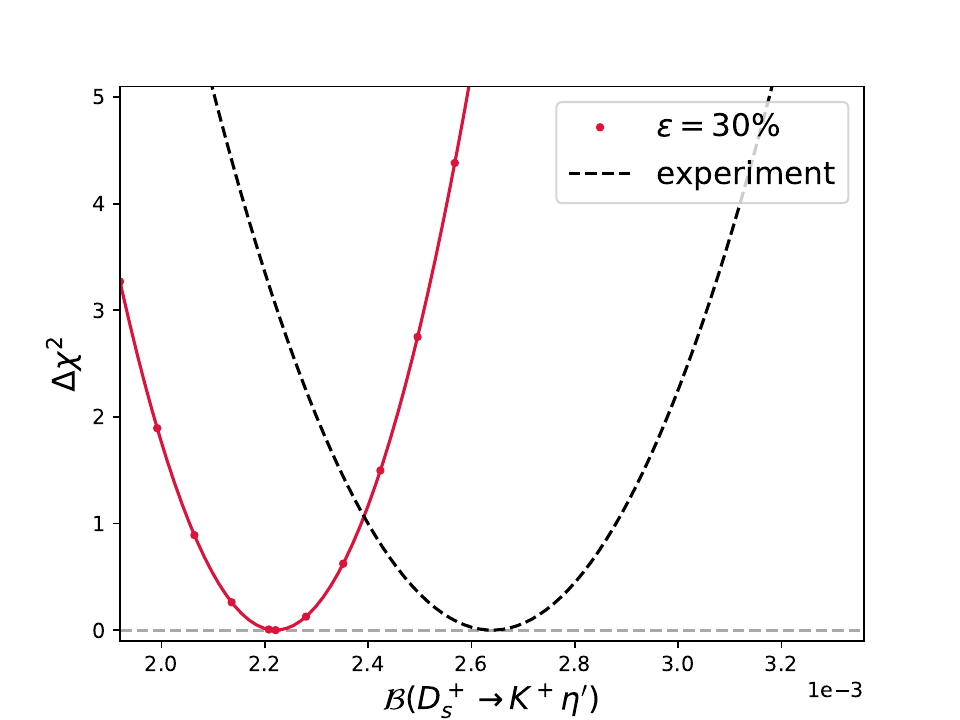}
  \end{minipage}
    \caption{Branching ratio scans for the global theory fit with $\varepsilon=30\%$~(red) compared to the experimental results~(black) for $D^+\to K^+\eta'$~(left) and $D_s^+\to K^+\eta'$~(right).}
    \label{fig:brscans}
\end{figure}

\begin{figure}[t]
  \includegraphics[width=\textwidth]{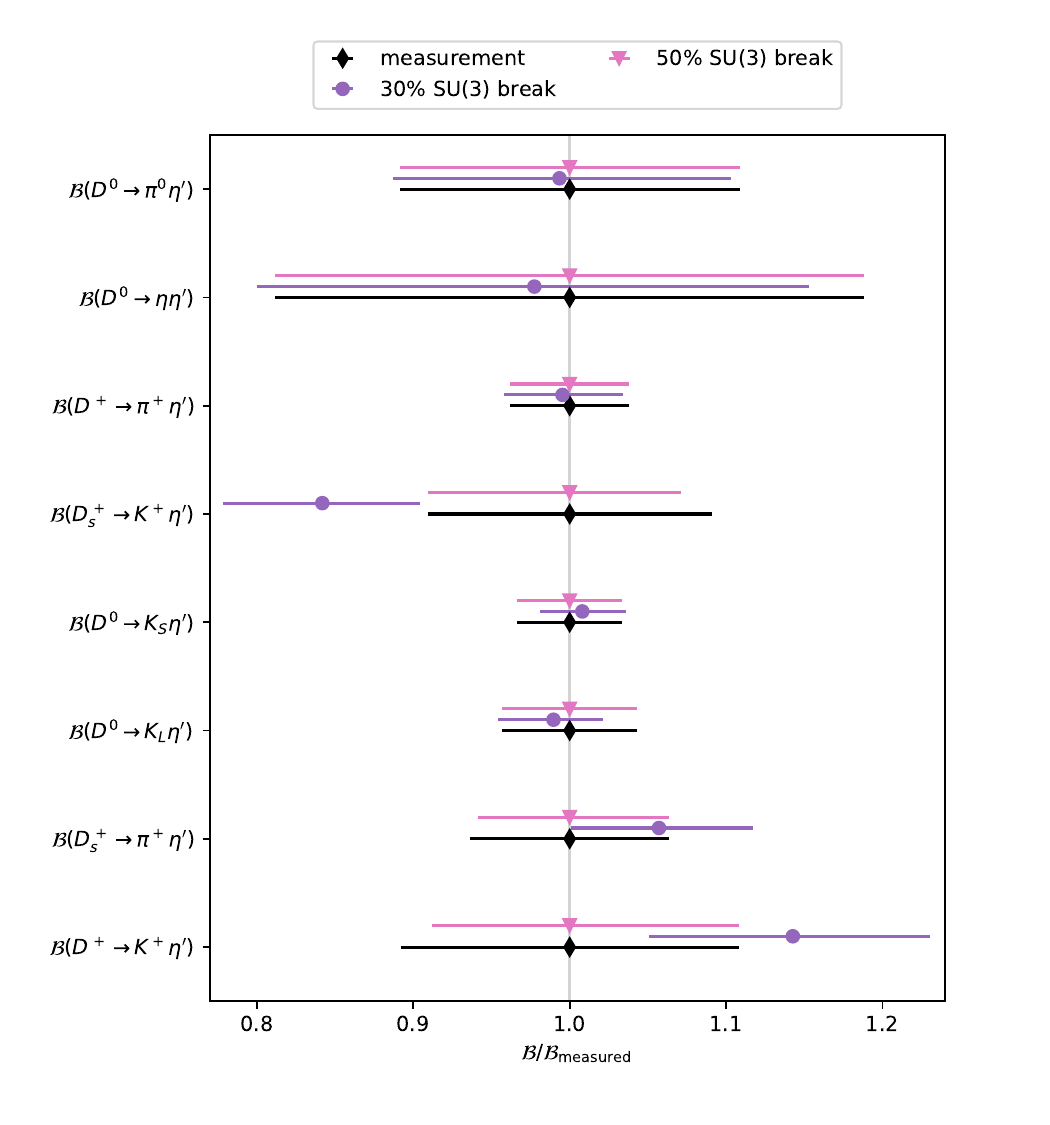}
  \caption{Comparison of experimental input data (black) with the results of our global fit allowing for 30\% SU(3)$_F$ breaking (purple) and 50\% SU(3)$_F$ breaking (pink). All branching ratios are normalised to the central value of the experimental results. }
  \label{fig:brplot}
\end{figure}

\subsection{Implications for branching ratios}

As already discussed, due to the multiple redefinitions of the underlying theory parameters, their interpretability is limited. We therefore concentrate here on the implications of our global fit for the branching ratios. To do so, we compute $\Delta\chi^2$ profiles of the branching ratios, comparing to the global minimum with a fixed value for $\varepsilon$.
The $1\sigma$ uncertainties are determined from the criterion $\Delta\chi^2=1$. Examples of these scans for our nominal $\varepsilon=30\%$ fit can be seen in Fig.~\ref{fig:brscans} for the decays $D_{(s)}^+\to K^+ \eta'$. 
A summary of all theoretical fit results are quoted in Table~\ref{tab:inputdata}. In Fig.~\ref{fig:brplot}, we compare our nominal fit results with the experimental data. All branching ratios are normalized to the central value of the experimental result. In addition, we also show the results from a fit with $50\%$ \SUt~breaking. In the latter case, we find excellent agreement with the experimental data and similar uncertainties ,~\emph{i.e.}~in this case we basically only reproduce the data. For our nominal fit, for $D_s^+\to K^+ \eta'$ we find a  $1.4\sigma$ deviation from the experimental result. Also $D^+\to K^+\eta'$ and $D_s^+\to \pi^+\eta'$ show a difference of 1.0 and 0.7$\sigma$, respectively. Improved branching ratio measurements of these decays would thus be useful to obtain more information on the amount of \SUt ~breaking. In addition, $D\to \pi^0\eta'$ and $D^0\to \eta\eta'$ are only known at the $10\%$ level, making updates of these modes also highly desirable. 

\subsection{Theory correlations between branching ratios}\label{sec:2d}

\begin{figure}[t]
  \begin{minipage}[c]{0.48\textwidth}
    \includegraphics[width=\textwidth]{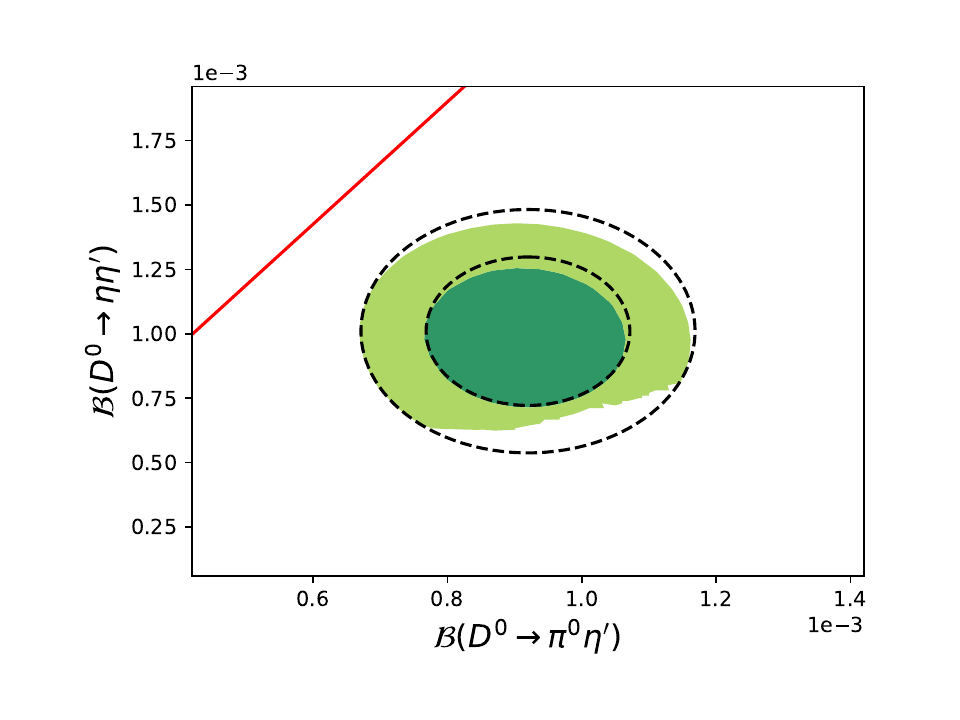}
  \end{minipage}
  \begin{minipage}[c]{0.48\textwidth}
    \includegraphics[width=\textwidth]{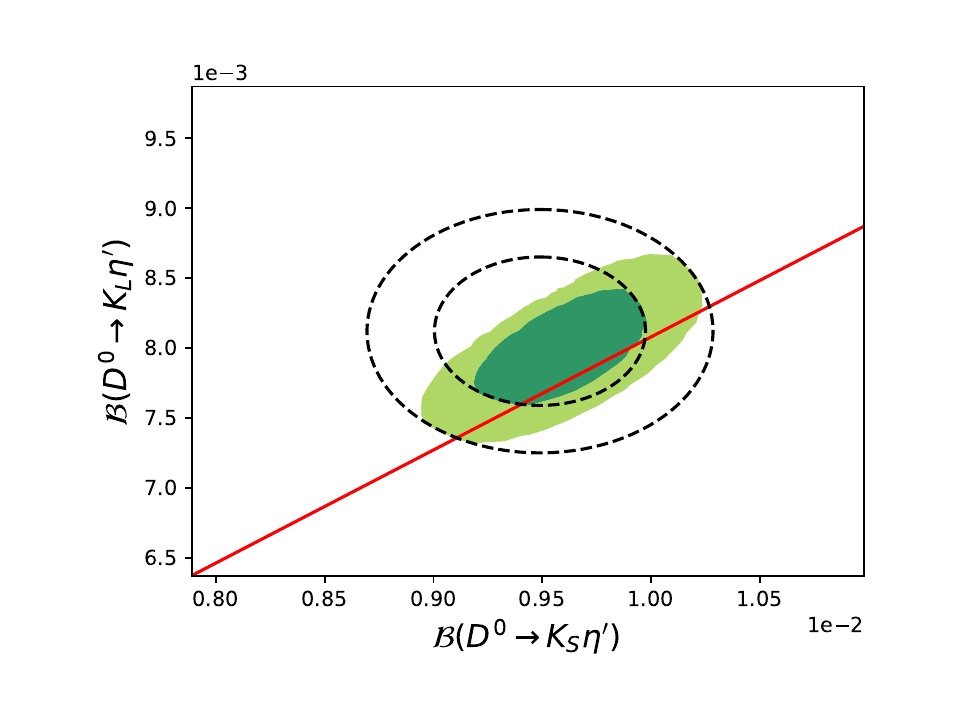}
  \end{minipage}
  \caption{Correlations between branching ratios of the neutral modes. The red line represents the \SUt{}  limit sum rule. The shaded areas correspond to the global fit result and the black lines to the experimental determinations. Contour lines represent the $68.30\%$ and $95.45\%$ confidence levels.}
    \label{fig:br2dneutral}
\end{figure}

\begin{figure}[t]
  \begin{minipage}[c]{0.48\textwidth}
    \includegraphics[width=\textwidth]{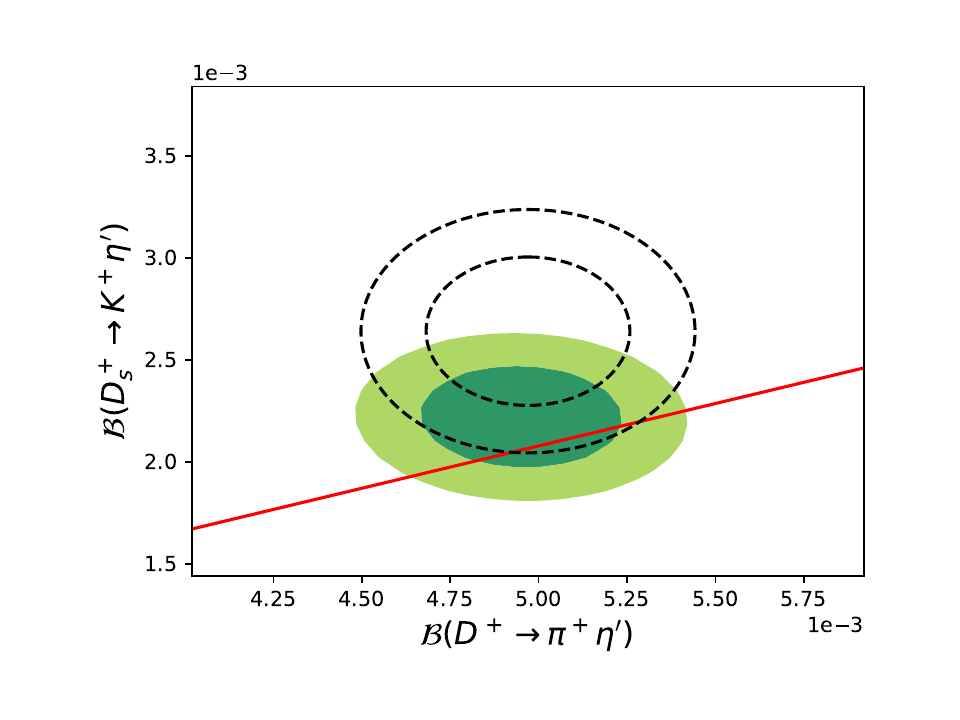}
  \end{minipage}\hfill
  \begin{minipage}[c]{0.48\textwidth}
    \includegraphics[width=\textwidth]{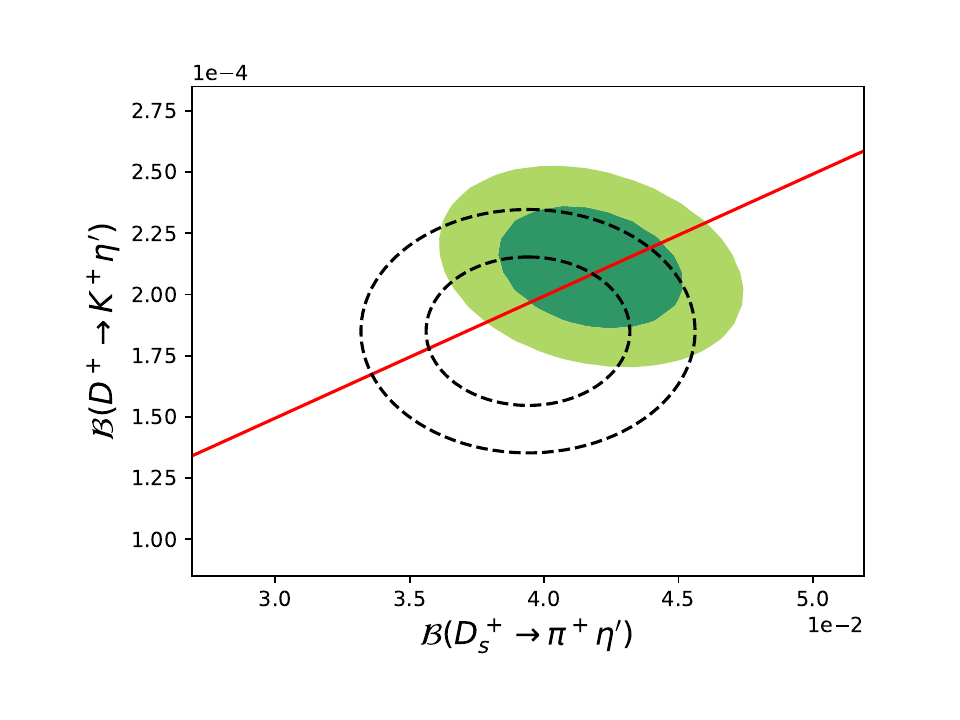}
  \end{minipage}
  \begin{minipage}[c]{0.48\textwidth}
    \includegraphics[width=\textwidth]{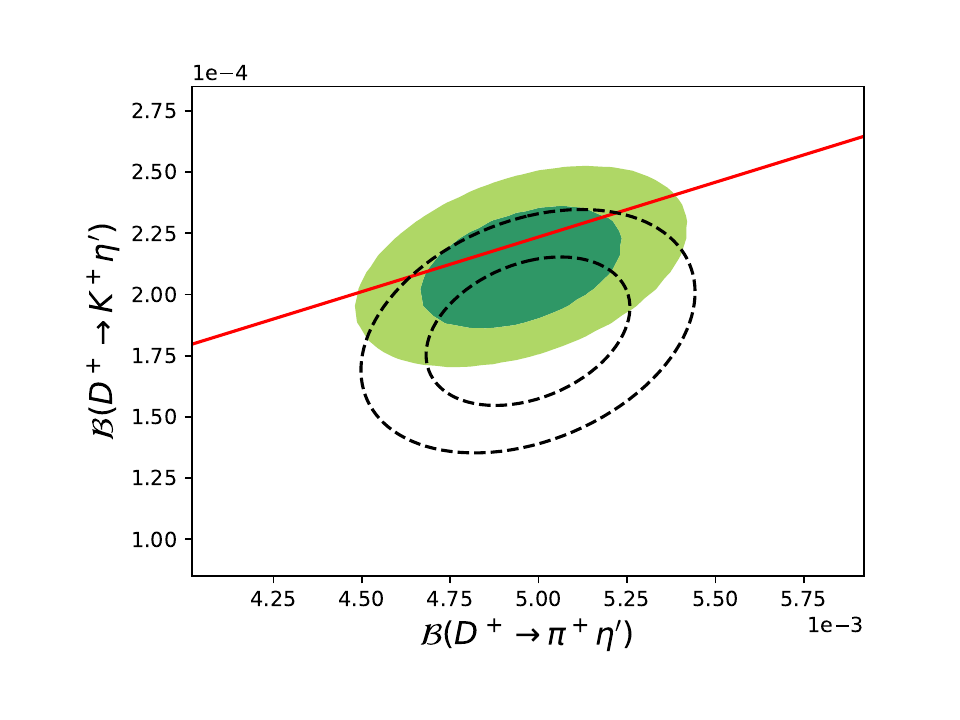}
  \end{minipage}
  \begin{minipage}[c]{0.48\textwidth}
    \includegraphics[width=\textwidth]{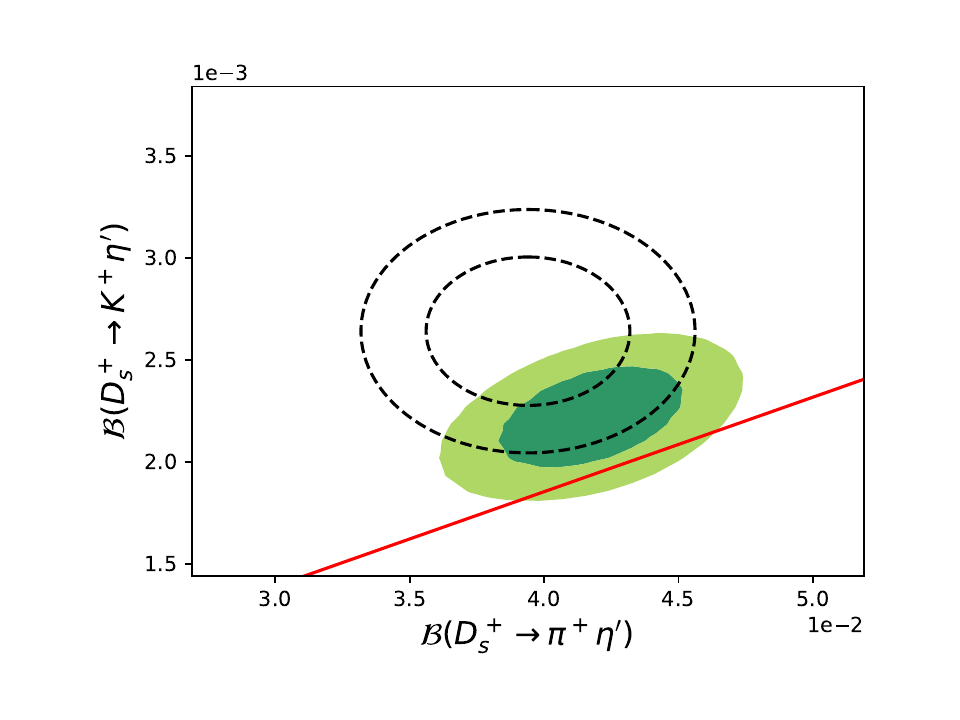}
  \end{minipage}
  \begin{minipage}[c]{0.48\textwidth}
    \includegraphics[width=\textwidth]{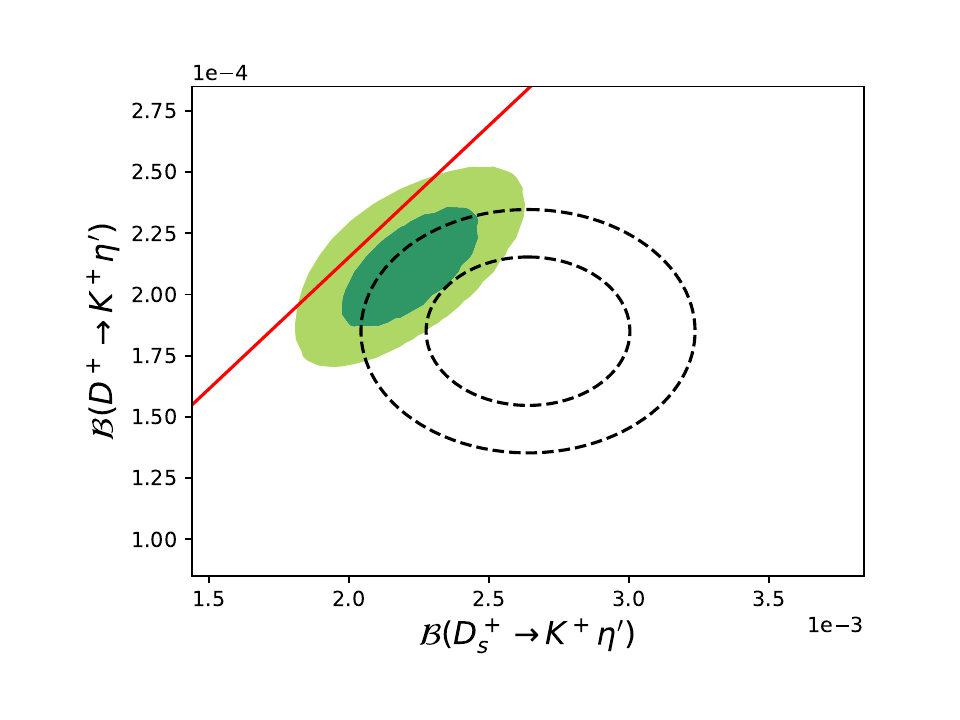}
  \end{minipage}
    \caption{Correlations between branching ratios for the charged modes. The red line represents the \SUt{}  limit sum rule. The shaded areas correspond to the global fit result and the black lines to the experimental determinations. Contour lines represent the $68.30\%$ and $95.45\%$ confidence levels.}
    \label{fig:br2d}
\end{figure}

Besides the global fit results for the individual branching ratios, it is interesting to consider correlations between different branching ratios stemming from the underlying SU(3)$_F$ symmetry. In Fig.~\ref{fig:br2dneutral} and ~\ref{fig:br2d}, we show the theory correlations between branching ratios of several neutral and charged decays, respectively. We consider the correlations between sets of charged and neutral decays because we have separate sum rules for these sets given in \eqref{eq:sr1} and \eqref{eq:sr2}. In addition, the neutral and charged modes are completely separate systems in the \SUt~limit. To highlight this, we also show the \SUt-limit relations between the respective decays. For completeness, we also show the experimental data given in Table~\ref{tab:inputdata}.

In Fig.~\ref{fig:br2dneutral}, we give the correlation between the SCS modes $(D^0\to \pi^0 \eta',D^0\to \eta\eta')$. Interestingly, we observe that these decays, while being in perfect agreement with the experimental data, differ significantly from the \SUt-limit prediction. For the modes with a neutral $K^0$, observed through their physical $K_S$ and $K_L$ states, we consider the correlation between 
$(D^0\to K_S \eta',D^0\to K_L\eta')$. 
In the \SUt-limit, the  branching ratios of these decays are given by 
\begin{align}
\mathcal{B}(D^{0}\to K_S \eta') \sim |\hat{C}_{81}|^2 (1 + 2 \lambda^2 ) + \mathcal{O}(\lambda^4)\,, \nonumber\\
\mathcal{B}(D^{0}\to K_L \eta') \sim |\hat{C}_{81}|^2 (1 - 2 \lambda^2 )+ \mathcal{O}(\lambda^4)\,,
\label{eq:dke}
\end{align}
because the physical states contain both CF and DCS parts. We observe that the data and our fit results in these modes are in agreement with \SUt-flavour symmetry. As expected, we also observe a strong theory correlation between these decays. It would therefore be interesting to also consider possible experimental correlations between these decays in future measurements.

For the charged decays, we consider several modes shown in Fig.~\ref{fig:br2d}. First, we show the set of SCS decays;  $(D^+\to \pi^+ \eta',D_s^+\to K^+\eta')$ and combine the CF and DCS modes; $(D_s^+\to \pi^+ \eta',D^+\to K^+\eta')$. Although these decays are described largely by the same parameters (see Table~\ref{tab:hatparam}), we do not observe strong correlations in our theory predictions.

For completeness, we also show the correlations between decays with the same initial $D^+$ and $D_s^+$ state: $(D^+\to \pi^+ \eta',D^+\to K^+\eta')$ and $(D_s^+\to \pi^+ \eta',D_s^+\to K^+\eta')$. For the latter, we notice also a clear deviation of the experimental measurements and the \SUt-limit correlation. Last, we show the correlation between $(D_s^+\to K^+ \eta',D^+\to K^+\eta')$, because these decays show the largest tension with the experimental results.

The two-dimensional correlations in Fig.~\ref{fig:br2dneutral} and Fig.~\ref{fig:br2d} clearly show how several sets of decays deviate (or agree with) the \SUt-limit sum rules. Therefore, it would be interesting to have more precise corresponding branching ratio measurements in the future including their correlations.

\section{Conclusions \label{sec:conclusions}}

We presented a first study of \SUt-breaking effects in $D\to P\eta'$ using a consistent treatment of the singlet-octet mixing between $\eta_0$ and $\eta_8$ within the SU(3)$_F$ power counting. Our universal mixing angle $\theta$ is not an observable quantity, unavoidably $\theta$ always appears together with the matrix elements.  As a consequence $D\to P\eta'$ decays become uncorrelated to $D\to P\eta$.
We find that the \SUt{} limit is ruled out by the data at 5.6$\sigma$ when compared to a fit with $50\%$ SU(3)$_F$ breaking, which gives a perfect description of the data. Allowing for $30\%$ \SUt{} breaking at the amplitude level, we show that the data can be consistently described. From the underlying theoretical expansion, we are able to predict the eight $D\to P\eta'$ branching ratios and the correlations between several decay channels. 

We find 1$\sigma$ differences between our theoretical prediction and measurements of the channels $D_s\to K^+\eta'$ and $D^+\to K^+ \eta'$. We predict that if \SUt-breaking is $30\%$ or smaller, future measurements of the branching ratio of $D_s\to K^+\eta'$ go down and of $D^+\to K^+ \eta'$ go up. If on the contrary, future measurements do not show this trend, we have to conclude that \SUt{} breaking is larger than expected.  As such, updated measurements of these decays are highly desired to shed further light on \SUt-breaking effects in the charm sector. We point out that especially the two-dimensional correlations between sets of charged or neutral decays can clearly show whether \SUt{} symmetry is respected (or broken) in subsets of decays. Doing so, we found that the channels $D^0\to \pi^0\eta'$ and $D^0\to \eta\eta'$ show the largest deviation from the \SUt{} limit. Clearly such decays are challenging to measure, but we emphasize that improved measurements of their branching ratios would allow to further test the quality of \SUt{} symmetry in the charm sector. 

Our thorough analysis of $D\to P\eta'$ serves as a first step towards predicting CP asymmetries in these modes. Finally, it would be interesting to compare our analysis with an updated analysis of \SUt{} breaking in the $D\to PP$ modes (including the $\eta$ channels) which is currently in progress \cite{inprogres}. 

\acknowledgments

S.S. is supported by a Stephen Hawking Fellowship from UKRI under reference EP/T01623X/1 and the STFC research grant ST/X00077X/1. The work of K.K.V. is
supported in part by the Dutch Research Council (NWO) as part of the project Solving
Beautiful Puzzles (VI.Vidi.223.083) of the research programme Vidi. U.N. acknowledges support by BMBF
grant 05H21VKKBA, \emph{Theoretische Studien f\"ur Belle II und LHCb}.

\appendix

\section{Notation}
\label{app:conv}

We employ the following sign convention
\begin{align}
\ket{K^+} &= \ket{u\bar{s}}\,, & 
\ket{K^0} &= \ket{d\bar{s}}\,, \\
\ket{K^-} &= \ket{s\bar{u}}\,, &
\ket{\overline{K}^0} &= \ket{s\bar{d}}\,,\\
\ket{\pi^+} &= \ket{u\bar{d}}\,, &
\ket{\pi^0} &= \frac{1}{\sqrt{2}} \left(\ket{u\bar{u}} - \ket{d\bar{d}}  \right)\,,\\
\ket{\pi^-} &= \ket{d\bar{u}}\,, &
\ket{D^0} &= \ket{c\bar{u}}\,,\\
\ket{D^+} &= \ket{c\bar{d}}\,, & 
\ket{D_s^+} &=\ket{c\bar{s}}\,,
\label{eq:convention}
\end{align}
found from  $M^a \sim (\bar u, \bar d,\bar s) \lambda^a (u,d,s)^T$ with the Gell-Mann  matrices $\lambda^a$ and $\ket{\pi^-} = \ket{M^-} = \ket{ M^1+ i M^2}/2$, $\ket{\pi^0} = \ket{M^3}/\sqrt{2}$, and so on.
Note that this sign convention deviates from the one used in other works~\cite{Muller:2015lua, Muller:2015rna, Gronau:1994rj}.
The meson octet thus corresponds to 
\begin{align}
\ket{\pi^+}, \ket{\pi^0}, \ket{\pi^-}, \ket{K^+}, \ket{K^0}, 
\ket{\bar K{}^0}, \ket{K^-}, \ket{\eta_8}.
\end{align}
For the $K_S$ and $K_L$ decays, we write the physical states as
\begin{align}
    \ket{K_S} = \frac{\ket{K^0} - \ket{{\overline{K}^0}}}{\sqrt{2}}, \quad\quad
    \ket{K_L} = \frac{\ket{K^0} + \ket{{\overline{K}^0}}}{\sqrt{2}} \ ,
    \label{eq:kskl}
\end{align}
which mixes the CF and DCS amplitudes. 

Here $C\ket{K^0} = \ket{\bar K^0}$, \textit{i.e.}\ $CP\ket{K^0} = -\ket{\bar K^0}$ is 
used. The sign convention for the charge conjugation operation $C$ is linked to the 
conventions in \eq{eq:convention}, because one can transform $\ket{K^0}$ into 
$\ket{\bar K^0}$ in two ways, by applying $C$ or by rotating the state around the $y$-axis in U-spin space with angle $\pi$, and both transformations must be compatible with each other \cite{Muller:2015lua}.
If one erroneously exchanges the "$-$" and "$+$" signs in \eq{eq:kskl}, one will find 
the wrong hierarchy among the branching fractions in \eq{eq:dke}. 
The combination of the mentioned U-spin rotation and $C$ is called $G_U$ parity \cite{Karliner:2010xb,Sahoo:2015msa,Meng:2022ozq}, 
which is the analogue of Lee's and Yang's $G$ parity based on isospin \cite{Lee:1956sw}. $\ket{K^0}$ and
$\ket{\bar K^0}$ are eigenstates of $G_U$ with eigenvalue $-1$; the requirement that all members of the U-spin triplet have the same $G_U$ quantum number fixes the sign convention for $C$ and implies the 
"$-$" sign in $\ket{K_S}$ in \eq{eq:kskl}.

\section{SU(3)$_F$-breaking topological diagrams and decomposition without redefinitions \label{app:su3break}}

\begin{figure}[t]
\begin{center}
\subfigure[\, $T_{88}^{(1)}$]{
        \includegraphics[width=0.2\textwidth]{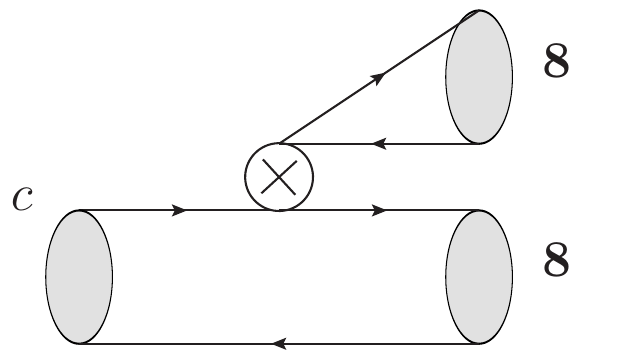}
}
\subfigure[\, $T_{ij,1}^{(1)}$]{
        \includegraphics[width=0.2\textwidth]{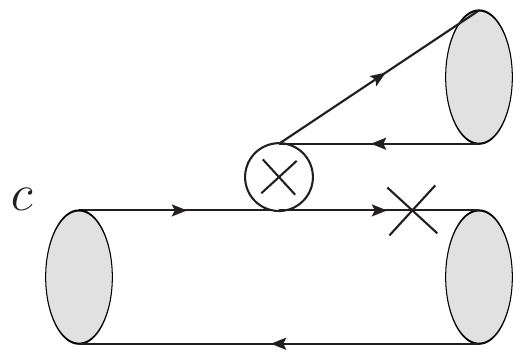}
}
\hfill %
\subfigure[\, $T_{ij,2}^{(1)}$]{
        \includegraphics[width=0.2\textwidth]{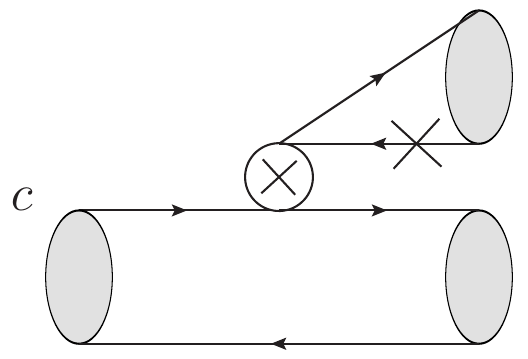}
}
\hfill %
\subfigure[\, $T_{ij,3}^{(1)}$]{
        \includegraphics[width=0.2\textwidth]{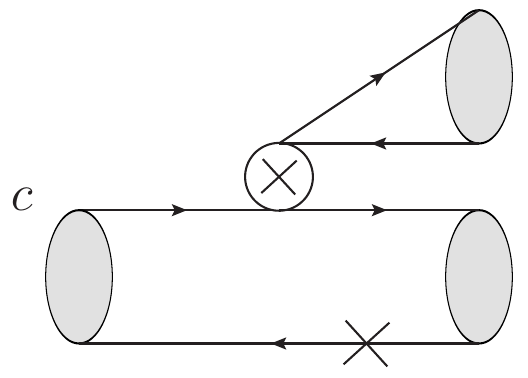}
}
\end{center}
\caption{
SU(3)$_F$-breaking tree topologies.
 \label{fig:feyn-diags-tree}}
\end{figure}

\begin{figure}[t]
\begin{center}
\subfigure[\, $A_{88}^{(1)}$]{
        \includegraphics[width=0.2\textwidth]{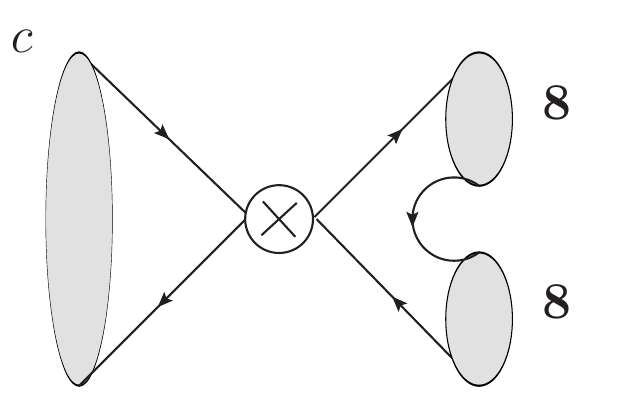}
}
\hfill
\subfigure[\, $A_{H,1}^{(1)}$]{
        \includegraphics[width=0.2\textwidth]{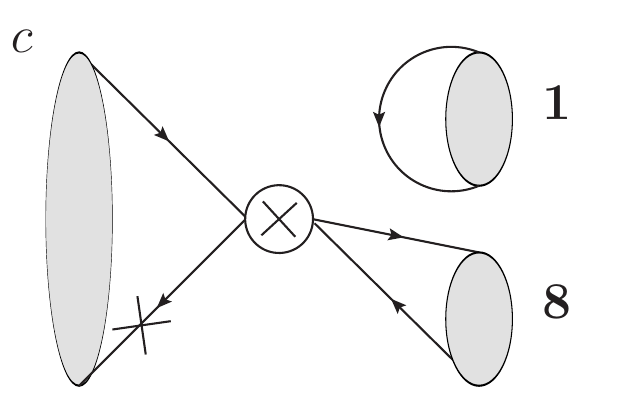}
}
\hfill
\subfigure[\, $A_{H,2}^{(1)}$]{
        \includegraphics[width=0.2\textwidth]{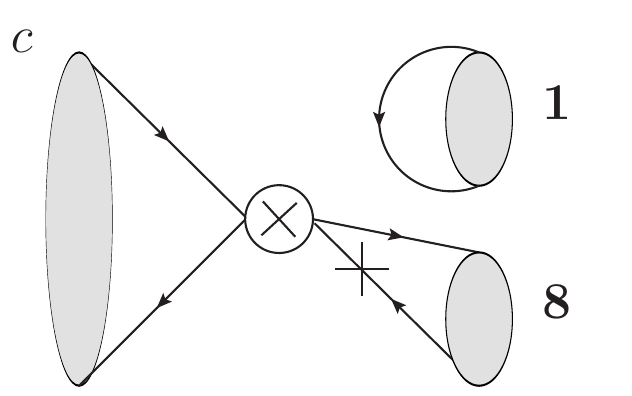}
}
\\
\subfigure[\, $A_{ij,1}^{(1)}$]{
        \includegraphics[width=0.2\textwidth]{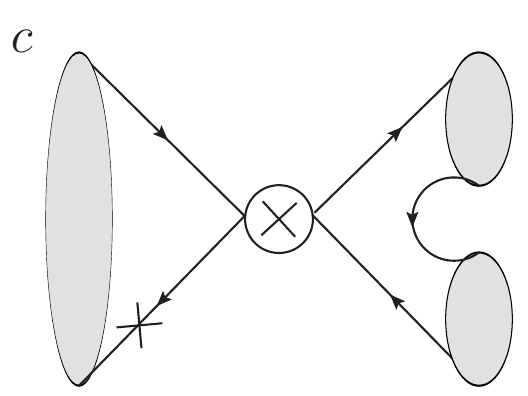}
}
\hfill %
\subfigure[\, $A_{ij,2}^{(1)}$]{
        \includegraphics[width=0.2\textwidth]{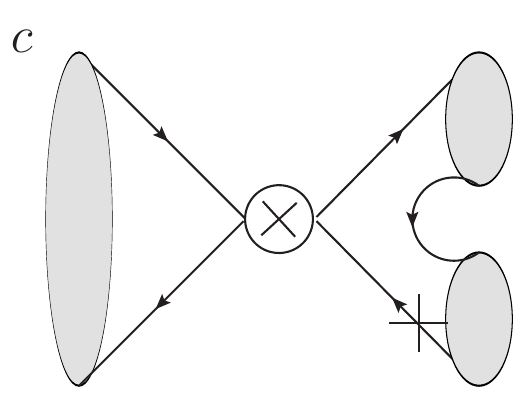}
}
\hfill %
\subfigure[\, $A_{ij,3}^{(1)}$]{
        \includegraphics[width=0.2\textwidth]{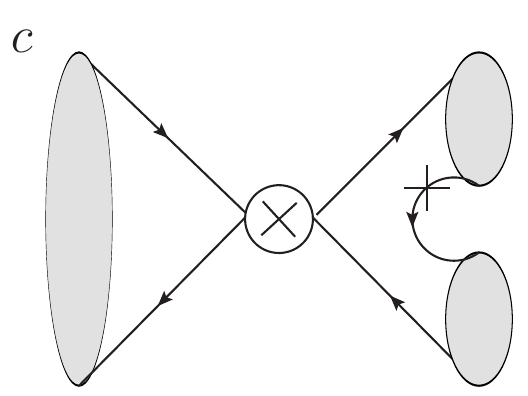}
}
\end{center}
\caption{
SU(3)$_F$-breaking annihilation topologies.
\label{fig:feyn-diags-annihilation}}
\end{figure}

\begin{figure}[t]
\begin{center}
\subfigure[\, $C_{88}^{(1)}$]{
        \includegraphics[width=0.2\textwidth]{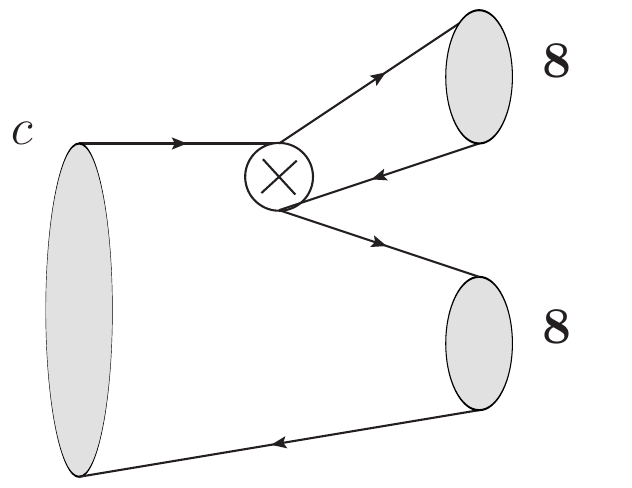}
}
\hfill %
\subfigure[\, $C_{ij,1}^{(1)}$]{
        \includegraphics[width=0.2\textwidth]{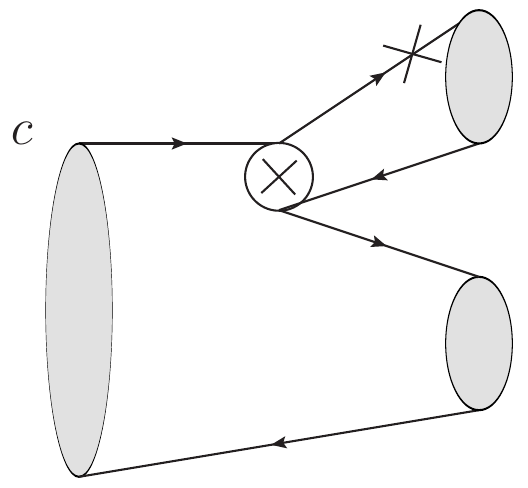}
}
\hfill %
\subfigure[\, $C_{ij,2}^{(1)}$]{
        \includegraphics[width=0.2\textwidth]{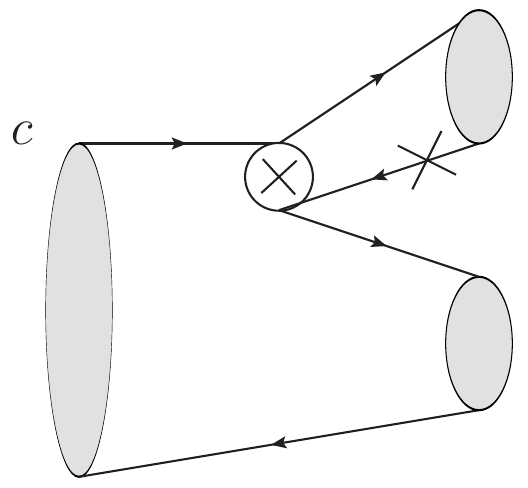}
}
\hfill %
\subfigure[\, $C_{ij,3}^{(1)}$]{
        \includegraphics[width=0.2\textwidth]{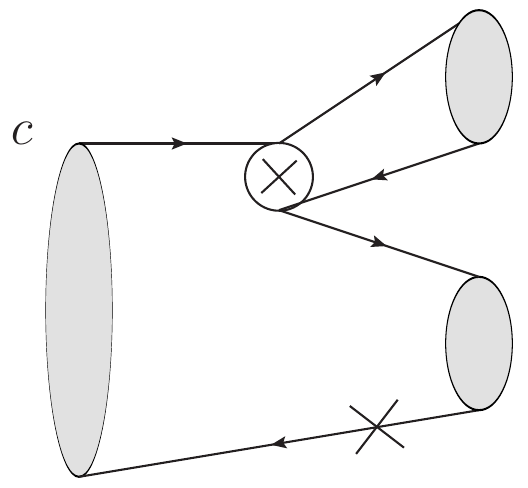}
}
\end{center}
\caption{
SU(3)$_F$-breaking color-suppressed topologies.
 \label{fig:feyn-diags-color-suppressed}}
\end{figure}

\begin{figure}[t]
\begin{center}
\subfigure[\, $E_{88}^{(1)}$]{
        \includegraphics[width=0.2\textwidth]{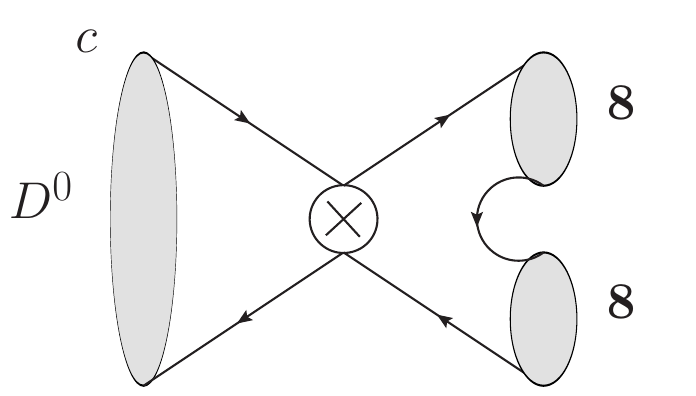}
}
\hfill %
\subfigure[\, $E_{ij,1}^{(1)}$]{
        \includegraphics[width=0.2\textwidth]{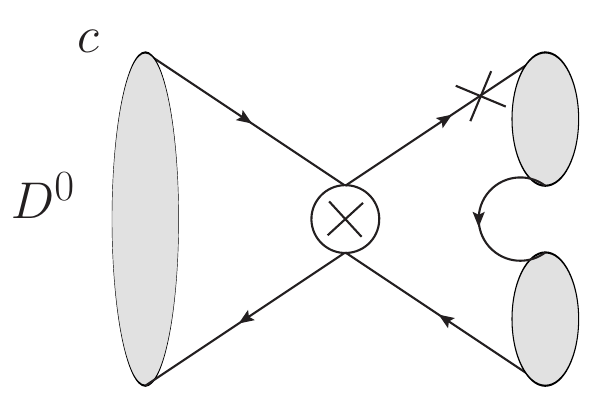}
}
\hfill %
\subfigure[\, $E_{ij,2}^{(1)}$]{
        \includegraphics[width=0.2\textwidth]{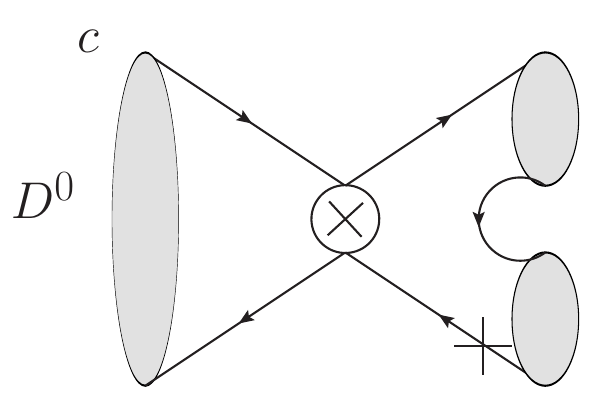}
}\\
\subfigure[\, $E_{ij,3}^{(1)}$]{
        \includegraphics[width=0.2\textwidth]{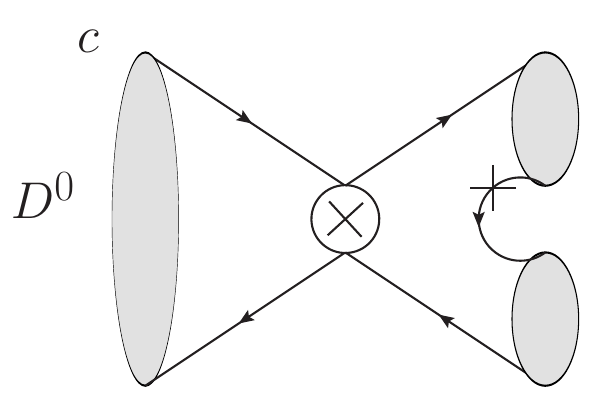}
}
\hfill
\subfigure[\, $E_{H,1}^{(1)}$]{
        \includegraphics[width=0.2\textwidth]{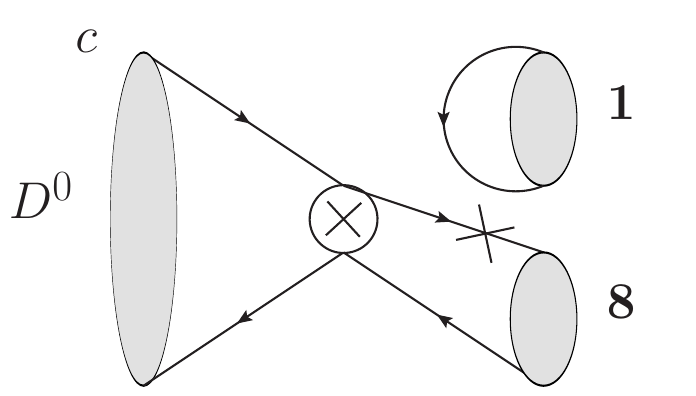}
}
\hfill
\subfigure[\, $E_{H,2}^{(1)}$]{
        \includegraphics[width=0.2\textwidth]{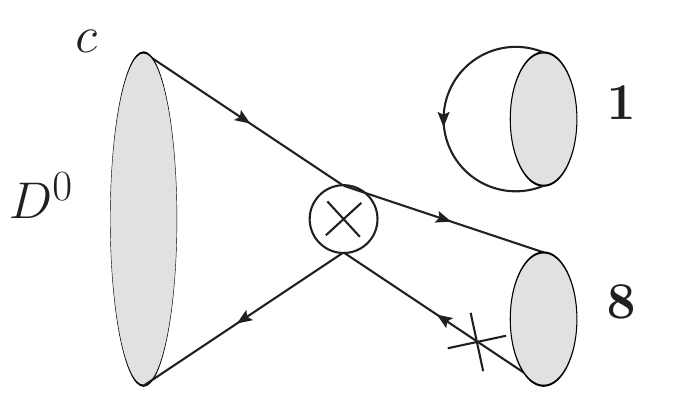}
}
\end{center}
\caption{
SU(3)$_F$-breaking exchange topologies.
The SU(3)$_F$-breaking effects within the hairpin itself can be absorbed into the SU(3)$_F$ limit hairpin diagram.
Note further that there are 3 SU(3)$_F$ breaking topologies for each of the SU(3)$_F$ limit exchange topologies,~\emph{i.e.},~a total of nine different SU(3)$_F$ breaking exchange diagrams plus two SU(3)$_F$ breaking hairpin diagrams. 
\label{fig:feyn-diags-exchange}}
\end{figure}

\begin{figure}[t]
\begin{center}
\subfigure[\, $P_{18, \mathrm{break}}^{(1)}$]{
        \includegraphics[width=0.3\textwidth]{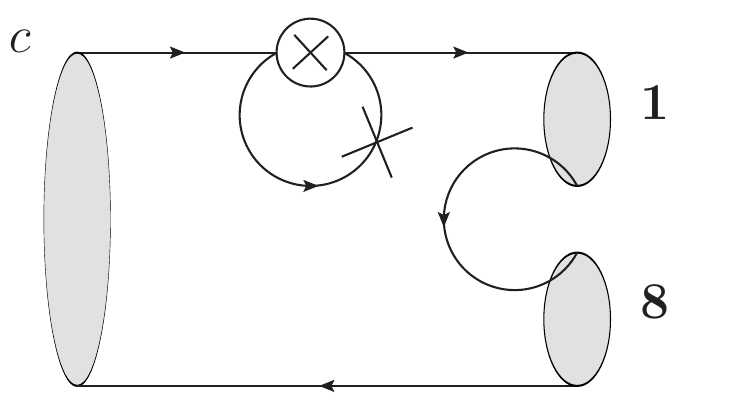}
}
\qquad %
\subfigure[\, $P_{81, \mathrm{break}}^{(1)}$]{
        \includegraphics[width=0.3\textwidth]{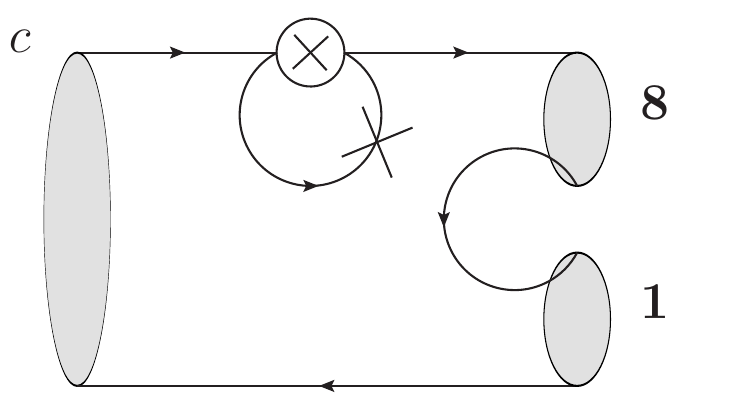}
}
\end{center}
\caption{
SU(3)$_F$-breaking penguin topologies.
 \label{fig:feyn-diags-broken-penguin}}
\end{figure}

In Figs.~\ref{fig:feyn-diags-tree}--\ref{fig:feyn-diags-broken-penguin} we show all topological Feynman diagrams with linear \SUt-breaking denoted with a superscript $(1)$. We show the SU(3)$_F$ decomposition without redefinitions in Table~\ref{tab:topoparametrizationprimesuppressed}.

\begin{landscape}
\begin{table}[t]
\begin{tiny}
\begin{center}
\setlength{\tabcolsep}{1pt} 
\begin{tabular} 
{c|c|c|c|c|c|c|c|c|c|c|c|c|c|c|c|c|c|c|c|c|c|c|c|c|c|c|c|c}
\hline \hline
Decay & 
 $T_{18,1}^{(1)}$ & $T_{18,2}^{(1)}$ & $T_{18,3}^{(1)}$ &  
 $T_{88}^{(1)}$ &
 $A_{18,1}^{(1)}$ & $A_{18,2}^{(1)}$ & 
 $A_{81,1}^{(1)}$ & $A_{81,2}^{(1)}$ & $A_{81,3}^{(1)}$ & 
 $A_{88}^{(1)}$ &
 $A_{H,1}^{(1)}$& $A_{H,2}^{(1)}$ & 
 $C_{18,1}^{(1)}$ & $C_{18,2}^{(1)}$ &  
 $C_{81,1}^{(1)}$ & $C_{81,2}^{(1)}$ & 
 $C_{88}^{(1)}$ &
 $E_{18,1}^{(1)}$ & $E_{18,2}^{(1)}$ & $E_{18,3}^{(1)}$ & 
 $E_{81,1}^{(1)}$ & $E_{81,2}^{(1)}$ & $E_{81,3}^{(1)}$ & 
 $E_{88}^{(1)}$ &
 $E_{H,1}^{(1)}$ & $E_{H,2}^{(1)}$ &
$P_{18,\mathrm{break}}^{(1)}$ &  $P_{81,\mathrm{break}}^{(1)}$ \\\hline\hline
\multicolumn{29}{c}{SCS} \\\hline\hline
$D^0 \rightarrow \pi^0 \eta'$  &  
$0$ & $0$ & $0$ &  
$0$ &
$0$ & $0$ &  
$0$ & $0$ & $0$ & 
$0$ &
  $0$ & $0$ &
 $\frac{1}{\sqrt{6}}$ & $\frac{1}{\sqrt{6}}$ & 
 $0$ & $0$ &
 $-\frac{1}{\sqrt{3}}$ &
 $0$ & $0$ & $0$ & 
 $0$ & $0$ & $0$ & 
 $\frac{1}{\sqrt{3}}$ &
 $0$ & $0$ &
$\frac{1}{\sqrt{6}}$ &  $\frac{1}{\sqrt{6}}$\\\hline\hline
$D^0 \rightarrow \eta \eta'$  &  
$0$ & $0$ & $0$ &  
$0$ &
$0$ &  $0$ &  
$0$ &  $0$ & $0$ & 
$0$ &
  $0$ & $0$ &
 $\frac{1}{3\sqrt{2}}$ & $\frac{1}{3\sqrt{2}}$ &  
 $-\frac{\sqrt{2}}{3}$ & $-\frac{\sqrt{2}}{3}$ & 
 $-1$ &
 $-\frac{\sqrt{2}}{3}$ & $-\frac{\sqrt{2}}{3}$ & $-\frac{\sqrt{2}}{3}$ & 
 $-\frac{\sqrt{2}}{3}$ & $-\frac{\sqrt{2}}{3}$ & $-\frac{\sqrt{2}}{3}$ & 
 $1$ &
 $-\sqrt{2}$ & $-\sqrt{2}$ &
 $\frac{1}{3\sqrt{2}} $  &  $\frac{1}{3\sqrt{2}}$ \\\hline\hline
$D^+ \rightarrow \pi^+ \eta'$ &  
$0$ & $0$ & $0$ &  
$-\frac{1}{\sqrt{6}}$ &
$0$ & $0$ &  
$0$ & $0$ & $0$ & 
$-\sqrt{\frac{2}{3}}$ &
  $0$ & $0$ &
 $\frac{1}{\sqrt{3}}$ & $\frac{1}{\sqrt{3}}$ &  
 $0$ & $0$ & 
 $-\sqrt{\frac{3}{2}}$ &
 $0$ & $0$ &  $0$ &
 $0$ & $0$ & $0$ & 
 $0$ &
 $0$ & $0$ &
$\frac{1}{\sqrt{3}}$ &  $\frac{1}{\sqrt{3}}$ \\\hline\hline
$D_s^+ \rightarrow K^+ \eta'$ &  
$\frac{1}{\sqrt{3}}$ & $\frac{1}{\sqrt{3}}$ & $\frac{1}{\sqrt{3}}$ & 
$-\sqrt{\frac{2}{3}}$ &
$\frac{1}{\sqrt{3}}$ &  $\frac{1}{\sqrt{3}}$ &  
$\frac{1}{\sqrt{3}}$ &  $\frac{1}{\sqrt{3}}$ & $\frac{1}{\sqrt{3}}$ & 
$-\frac{1}{\sqrt{6}}$ &
 $\sqrt{3}$ &  $\sqrt{3}$ & 
 $\frac{1}{\sqrt{3}}$ & $\frac{1}{\sqrt{3}}$ &  
 $0$ & $0$ & 
 $-\sqrt{\frac{3}{2}}$ &
 $0$ & $0$ & $0$ & 
 $0$ & $0$ & $0$ & 
 $0$ &
 $0$ & $0$ &
$\frac{1}{\sqrt{3}}$ &  $\frac{1}{\sqrt{3}}$ \\\hline\hline

\multicolumn{29}{c}{CF} \\\hline\hline
$D^0 \rightarrow \overline{K}^0 \eta'$ &  
 $0$ & $0$ & $0$ &  
 $0$ &
 $0$ &  $0$ &  
 $0$ &  $0$ & $0$ & 
 $0$ &
 $0$ & $0$ &
 $0$ & $0$ & 
 $\sqrt{\frac{1}{3}}$ & $0$ & 
 $\frac{1}{\sqrt{6}}$ &
 $\sqrt{\frac{1}{3}}$ & $0$ & $\sqrt{\frac{1}{3}}$ & 
 $\sqrt{\frac{1}{3}}$ & $0$ & $0$ & 
 $-\frac{1}{\sqrt{6}}$ &
 $\sqrt{3}$ & $0$ &
$0$ &  $0$ \\\hline\hline

$D_s^+ \rightarrow \pi^+ \eta'$  &   
$\frac{1}{\sqrt{3}}$ & $0$ & $\frac{1}{\sqrt{3}}$ &  
$-\sqrt{\frac{2}{3}}$ &
$\frac{1}{\sqrt{3}}$ & $0$ &  
$\frac{1}{\sqrt{3}}$ & $0$ & $0$ & 
$\sqrt{\frac{2}{3}}$ &
$\sqrt{3}$ &  $0$ & 
 $0$ & $0$ & 
 $0$ & $0$ & 
 $0$ &
 $0$ & $0$ & $0$ & 
 $0$ & $0$ & $0$ & 
 $0$ &
 $0$ & $0$ &
$0$ &  $0$ \\\hline\hline

\multicolumn{29}{c}{DCS} \\\hline\hline

$D^0 \rightarrow K^0 \eta'$ &  
$0$ & $0$ & $0$ &  
$0$ &
$0$ & $0$ &  
$0$ & $0$ & $0$ & 
$0$ &
 $0$ & $0$ &
 $0$ & $0$ &  
 $0$ & $\frac{1}{\sqrt{3}}$ & 
 $\frac{1}{\sqrt{6}}$ &
 $0$ & $\frac{1}{\sqrt{3}}$ & $0$ & 
 $0$ & $\frac{1}{\sqrt{3}}$ & $\frac{1}{\sqrt{3}}$ & 
 $-\frac{1}{\sqrt{6}}$ &
 $0$ & $\sqrt{3}$ &
$0$ &  $0$ \\\hline\hline

$D^+ \rightarrow K^+ \eta'$  &  
$0$ & $\frac{1}{\sqrt{3}}$ & $0$ &
$\frac{1}{\sqrt{6}}$ &
$0$ & $\frac{1}{\sqrt{3}}$ &  
$0$ & $\frac{1}{\sqrt{3}}$ & $\frac{1}{\sqrt{3}}$ & 
$-\frac{1}{\sqrt{6}}$ &
 $0$ & $\sqrt{3}$ & 
 $0$ & $0$ & 
 $0$ & $0$ & 
 $0$ &
 $0$ & $0$ & $0$ & 
 $0$ & $0$ & $0$ & 
 $0$ &
 $0$ & $0$ &
$0$ &  $0$ \\\hline\hline

\end{tabular}
\caption{SU(3)$_F$ breaking diagrams. Note that the SU(3)$_F$ limit diagram 88 is power suppressed for $\eta'$.  Note also that the diagrams $A_{18,3}^{(1)}$, $C_{18,3}^{(1)}$, $C_{81,3}^{(1)}$ do not contribute.} 
\label{tab:topoparametrizationprimesuppressed}
\end{center}
\end{tiny}
\end{table}
\end{landscape}

\clearpage

\bibliography{draft.bib}
\bibliographystyle{JHEP}

\end{document}